\documentclass{ws-ijmpd}
\usepackage[super,compress]{cite}
\usepackage{amsmath}
\usepackage{amssymb}
\usepackage{amsfonts}
\usepackage[utf8]{inputenc}
\usepackage[T1]{fontenc}
\usepackage{mathrsfs}
\usepackage{anyfontsize}
\usepackage[utf8]{inputenc} 
\DeclareUnicodeCharacter{00A0}{ }

\mathchardef\arr="017E 
\renewcommand\vec[1]{\setbox0=\hbox{$#1$}\lower2ex\hbox to 0pt{\hbox to \wd0{\hss$\arr\;$\hss}\hss}\box0}

\usepackage{pifont}

\usepackage[linktocpage,breaklinks]{hyperref}

\newcommand{\rs}{r_{\rm s}}
\newcommand{\dd}{\mathrm{d}}
\newcommand{\ccc}{{_{\rm c}}}
\newcommand{\mpl}{M_{\rm pl}}
\newcommand{\gsim}{\;\mbox{\raisebox{-0.5ex}{$\stackrel{>}{\scriptstyle{\sim}}$}
}\;}
\newcommand{\lsim}{\;\mbox{\raisebox{-0.5ex}{$\stackrel{<}{\scriptstyle{\sim}}$}
}\;}
\newcommand{\uo}{\Upsilon_1}
\newcommand{\ut}{\Upsilon_2}
\newcommand{\pn}{\Psi}
\newcommand{\ext}{{\rm ext}}

\newcommand{\nm}{\mu\nu}
\begin{document}

\markboth{Jeremy Sakstein}
{Astrophysical Tests of Screened MG}

%
\catchline{}{}{}{}{}
%

\title{Astrophysical Tests of Screened Modified Gravity}

\author{Jeremy Sakstein}

\address{Center for Particle Cosmology,
Department of Physics and Astronomy,
University of Pennsylvania,
209 S. 33rd St., Philadelphia, PA 19104, USA\\
sakstein@sas.upenn.edu}

\maketitle


%
%


\section{Introduction}

Screened modified gravity theories evade the solar system tests that have proved prohibitive for classical alternative gravity theories such as Brans-Dicke. In many cases, they do not fit into the PPN formalism. The environmental dependence of the screening has motivated a concerted effort to find new and novel probes of gravity using objects that are well-studied but have hitherto not been used to test gravity. Astrophysical objects---stars, galaxies, clusters---have proved competitive tools for this purpose since they occupy the partially-screened regime between solar system and the Hubble flow. In this section, we review the current astrophysical tests of screened modified gravity theories. We begin by introducing the theories we will study and outline the strategy typically employed to identify astrophysical probes. 

\subsection{Searching for Screening Mechanisms}
\label{sec:wheretolook}

We will split the known theories with screening mechanism into three distinct categories that exhibit similar effects on astrophysical objects. This allows us to identify the optimum strategy for testing each theory.

{\bf Thin-Shell Theories}: Chameleon \cite{Khoury:2003aq,Khoury:2003rn}, symmetron \cite{Hinterbichler:2010es}, and dilaton \cite{Brax:2010gi} models all screen using the thin-shell effect. For this reason we will refer to them as \emph{thin-shell} theories. The specific details of each model are not important for astrophysical tests and one can completely parameterize them using the effective coupling $\beta(\phi_{\rm BG})$ where $\phi_{\rm BG}$ is the asymptotic (background) field value and the self-screening parameter
\begin{equation}
\chi_{\rm BG}= \frac{\phi_{\rm BG}}{2\beta(\phi_{\rm BG})\mpl}.
\end{equation}
For $f(R)$ models one has $f_{R0}=2\chi_0/3$ where $\chi_0$ is the value of $\chi_{\rm BG}$ evaluated at cosmic densities. If the self-screening parameter is larger than an object's Newtonian potential $\pn=GM/R$ then this object will be self-screening. If not, then an object will be partially unscreened. This implies that the best objects for testing these theories are non-relativistic ones. In particular, main-sequence stars have $\pn\sim 10^{-6}$ whereas post-main-sequence stars have $\pn\sim 10^{-7}$--$10^{-8}$ (owing to their larger radii) and are therefore more constraining probes. Similarly, rotationally-supported galaxies have
\begin{equation}
\pn\sim \frac{GM}{R}= {v_{\rm circ}^2},
\end{equation}
where $v_{\rm circ}$ is the circular velocity. The most unscreened galaxies are therefore dwarf galaxies with $v_{\rm circ}\sim 50$ km/s so that $\pn\sim 10^{-8}$. (Spirals like the Milky Way have $v_{\rm circ}\sim 200$ km/s implying $\pn\sim 10^{-6}$.) There is the added complication of environmental screening whereby a potentially unscreened dwarf could be screened by its cluster companions. Therefore, one needs to use void dwarfs as laboratories for testing thin-shell screening theories. Reference \cite{Cabre:2012tq} has complied a `screening map' of the nearby universe using criteria developed by \cite{Zhao:2011cu} and calibrating on N-body simulations. Recently, this has been revisited by \cite{Desmond:2017ctk}.

{\bf Vainshtein Screening Theories}: Theories that screen using the Vainshtein mechanism so that the ratio of the scalar to Newtonian force outside a spherical object is
\begin{equation}
\frac{F_\phi}{F_{\rm N}}=2\alpha^2\left(\frac{r}{r_{\rm V}}\right)^n
\end{equation}
will be referred to as \emph{Vainshtein screening theories}. We parameterize the coupling strength $\alpha$ and Vainshtein radius (below which the force is screened) using a crossover scale $r_c(=\Lambda_{\rm c}^{-1})$ akin to the DGP model so that $r_{\rm V}^3\sim \alpha GMr_c^2$. These theories include very general theories including Horndeski \cite{Horndeski:1974wa} but here we will only focus on cubic and quartic galileon models \cite{Nicolis:2008in} for which $n=3/2$ and $n=2$ respectively. In the case of Vainshtein screening, the Vainshtein radius is typically larger than the radius of stars and galaxies making astrophysical tests difficult but not impossible. 

{\bf Vainshtein Breaking Theories}: Theories such as beyond Horndeski \cite{Gleyzes:2014dya,Gleyzes:2014qga} and degenerate higher-order scalar-tensor theories (DHOST) (see \cite{Langlois:2017mdk} and references therein) exhibit a `breaking of the Vainshtein mechanism' such that the Newtonian potential and lensing potential ($g_{ij}=(1-2\Phi)\delta_{ij}$) are corrected inside extended objects to \cite{Kobayashi:2014ida,Koyama:2015oma,Crisostomi:2017lbg} 
\begin{align}
\frac{\dd\pn}{\dd r}&=\frac{GM(r)}{r^2}+\frac{\uo G}{4}\frac{\dd ^2 M(r)}{\dd r^2}\\
\frac{\dd\Phi}{\dd r} & = \frac{GM(r)}{r^2}-\frac{5\ut G}{4r}\frac{\dd M(r)}{\dd r}+{\Upsilon_3 G}\frac{\dd ^2 M(r)}{\dd r^2}
\end{align}
where the three dimensionless parameters $\Upsilon_i$ are related to the cosmological values of the functions and parameters appearing in a specific theory and also the effective description of dark energy\cite{Sakstein:2017xjx,Crisostomi:2017lbg,Langlois:2017dyl,Dima:2017pwp} (introduced by \cite{Bellini:2014fua}). The form of the corrections do not suggest the best objects for testing these theories and one must calculate on an object by object basis.

\subsection{Roadmap of Astrophysical Tests}
\label{sec:roadmap_astro}

We will begin by discussing the most important difference between Vainshtein and thin-shell screened theories in section \ref{sec:EP}: equivalence principle violations. Next, we will introduce the theory of stellar structure in modified gravity in section \ref{sec:SS}. In the subsequent sections we review current astrophysical bounds by object. Non-relativistic stars in section \ref{sec:SStests}, galactic tests in section \ref{sec:galtests}, galaxy cluster tests in section \ref{sec:clustertests}, and tests using relativistic stars in section \ref{sec:relSS}. 

%
%

\section{Equivalence Principle Violations}
\label{sec:EP}

\subsection{Weak Equivalence Principle}

One important difference between thin-shell and Vainshtein screening is the presence of weak equivalence principle (WEP) violations\footnote{We define the WEP as the statement that the motion of a test body in an external gravitational field depends only on its mass and is independent of its composition and internal structure.}. It was pointed out in the original chameleon and symmetron papers that thin-shell screening violates the WEP \cite{Khoury:2003aq,Khoury:2003rn,Hinterbichler:2010es} because the thin-shell factors for each body, which determines their motion in an external field, are composition and structurally dependent. This issue was studied in more detail by \cite{Hui:2009kc} who also studied equivalence principle violations in galileon theories. Consider an extended object of inertial mass $M_i$ and gravitational mass $M_{g}$ in an applied external Newtonian potential $\pn^\ext$ and scalar field $\phi^\ext$ (chameleon, symmetron, or galileon). The equation of motion for this object in the non-relativistic limit is
\begin{equation}\label{eq:WEP1}
M_i\ddot{x}=-M_g\nabla\pn^\ext-\frac{Q}{\mpl}\nabla\phi^{\rm ext}.
\end{equation}
The gravitational mass can be thought of as a `gravitation charge' that parameterizes the response of the object to an externally applied Newtonian potential and so we have defined an analogous scalar charge $Q$ that quantifies the response of an object to an externally applied scalar field\footnote{The factor of $\mpl$ is needed because $\phi^\ext$ has different units to $\pn^\ext$. It is chosen so that that $Q$ has units of mass.}. In theories without a scalar field, the WEP is obeyed if $M_i=M_g$. This is not generically the case in scalar field theories because $Q$ can depend on the structure and composition of the object. In this section we will refer to the baryonic mass $M$ defined by
\begin{equation}\label{eq:baryonicmass}
M=\int\dd^3\vec{x} T^\mu_\mu=\int\dd^3\vec{x} \rho(x),
\end{equation}
where the second equality holds in the non-relativistic limit. Note that this may include the mass of dark matter but not the self-energy of the gravitational field, which is found by integrating the Landau-Lifschitz energy-momentum pseudo-tensor.

{\bf Thin-shell screening}: For theories that screen using the thin-shell effect (chameleon and symmetron theories) one has
\begin{equation}\label{eq:Qcham}
Q=\beta(\phi_{\rm BG})M\left(1-\frac{M(\rs)}{M}\right),
\end{equation}
where $M(\rs)$ is the mass enclosed inside the screening radius. The force between two bodies with masses $M_1$ and $M_2$ is \cite{Mota:2006ed,Mota:2006fz}
\begin{equation}
F_{1,\,2}=2Q_1Q_2\frac{GM_1M_2}{r^2},
\end{equation}
where $Q_i$ is given by equation \eqref{eq:Qcham} with $M\rightarrow M_i$. Thus the WEP is violated unless either $Q=0$ or $Q=M$ i.e. the objects are fully screened or fully unscreened. 

{\bf Vainshtein screening}: In the case of Vainshtein screening, there is no thin shell suppression. Furthermore, the equation of motion can be written in the form of a current conservation law $\nabla_\mu J^\mu=8\pi\alpha G \rho$, which ensures that
\begin{equation}\label{eq:QVain}
Q=M
\end{equation}
i.e. the charge is equal to the baryonic mass. The WEP is therefore satisfied in galileon theories. One possible caveat to this is many-body effects. The equation of motion for galileon theories is non-linear in second-derivatives, which leads to severe violations of the superposition principle\footnote{The equations of motion for chameleon and symmetron theories are quasi-linear and so there is always some regime in which superposition approximately holds. }. The above argument circumvents this by assuming that the external galileon field is only slowly varying so that the galilean shift symmetry can be used to superimpose the fields, and it is not clear what happens away from this approximation. The full two-body problem has been studied by \cite{Hiramatsu:2012xj} for an Earth-Moon like system; they found a mass-dependent reduction of the galileon force of $\sim 4\%$ indicating that the WEP may be broken by non-linear many-body effects. The non-linear nature of the equations makes modeling of such systems difficult. Indeed, departures from spherical symmetry do not have analytic solutions except in highly symmetric cases \cite{Bloomfield:2014zfa,Brito:2014ifa}. See \cite{Andrews:2013qva,Avilez-Lopez:2015dja} for some detailed studies of this issue.

\subsection{Strong Equivalence Principle}
\label{sec:SEP}

The strong equivalence principle (SEP) is the statement that an object's motion is independent of its self-gravity. Unlike the WEP, the SEP is violated by all of the theories considered in this section\footnote{In fact, scalar-tensor theories generically violate the SEP, the statements made in this subsection have nothing to do with screening.}. This is because the scalar field is sourced only by the baryonic mass (defined in \eqref{eq:baryonicmass}) and not the curvature so that the no-hair theorems hold and strongly gravitating objects have no scalar charge \footnote{One exception to this is scalar couplings to the Gauss-Bonnet scalar \cite{Kanti:1995vq,Sotiriou:2013qea,Sotiriou:2014pfa,Dong:2017toi,Silva:2017uqg}.}. A no-hair theorem for the galileon was proved by \cite{Hui:2012qt} for static, spherically symmetric black holes and subsequently generalized by \cite{Sotiriou:2013qea} to the case of slow rotation. Thus, if a system composed of baryonic matter (including dark matter) and black holes the baryonic component will have $Q=M$ while the black holes will have $Q=0$. The baryons will therefore fall at a faster rate than the black holes in an externally applied gravitational field, violating the SEP. In the case of chameleon theories, the presence of an accretion disk around black holes may source secondary scalar hair \cite{Davis:2014tea,Davis:2016avf}.

\section{Stellar Structure in Modified Gravity}	

\label{sec:SS}

Stars are complicated objects whose lives, existence, and stability are a result of the interface between diverse and disparate areas of physics, including gravitational physics, atomic physics, nuclear physics, hydrodynamics, and particle physics \cite{Kippenhahn1994}. Modern theoretical modeling of stellar structure and evolution therefore utilizes sophisticated numerical simulations that solve a large number (often in the thousands depending on the type of star) of coupled differential equations simultaneously. Fortuitously, the effects of gravitational physics appears in a single equation, the momentum equation, which describes the Lagrangian velocity $\vec{v}=\dot{\vec{r}}$ of a fluid element located at Lagrangian position $\vec{r}$ due to some external force (per unit mass) $\vec{f}$ and the hydrodynamic (Eulerian) pressure $P$:
\begin{equation}\label{eq:momentum_equation}
\dot{\vec{v}}=-\frac{1}{\rho}\nabla P + \vec{f},
\end{equation}
where $\rho(\vec{r})$ is the Eulerian density. In the case of general relativity, the force per unit mass is simply the gradient of the Newtonian potential
\begin{equation}
\vec{f}=-\nabla\pn.
\end{equation}
For alternative theories, one must solve for the force per unit mass within the new framework. Typically this involves solving for additional scalar (or other spin) field profiles sourced by the star's mass. Note that we will only discuss non-relativistic objects here, postponing relativistic stars for a later section.

\subsection{Equilibrium Structure}

The velocity of each fluid element is constant for a static, spherically symmetric object in equilibrium and so the left hand side of equation \eqref{eq:momentum_equation} is zero. In GR, the force per unit mass is simply the Newtonian force and one has the well-known hydrostatic equilibrium equation (HSEE)
\begin{equation}\label{eq:HSEGR}
\frac{\dd P(r)}{\dd r}=-\frac{G M(r)\rho(r)}{r^2},
\end{equation}
where $M(r)$ is the mass enclosed inside $r$ and therefore satisfies the continuity equation
\begin{equation}\label{eq:contstellar}
\frac{\dd M(r)}{\dd r^2}=4\pi r^2\rho(r).
\end{equation}
For thin-shell screening theories, the HSEE is modified to\footnote{Note that we have ignored the mass of the field, which is a good approximation inside the unscreened region of stars \cite{Chang:2010xh,Davis:2011qf,Jain:2012tn,Sakstein:2013pda,Sakstein:2015oqa}.} 
\begin{equation}\label{eq:HSECHAM}
\frac{\dd P(r)}{\dd r}=-\frac{G M(r)\rho(r)}{r^2}\left[1+2\beta^2(\phi_{\rm BG})\left(1-\frac{M(r_{\rm s})}{M(r)}\right)\Theta\left(r-\rs\right)\right],
\end{equation}
where $M(r_{\rm s})$ is the mass enclosed within the screening radius $\rs$, $\Theta(x)$ is the Heavyside step function, and the new factor arises from the fifth-force $F_\phi=-\beta(\phi_{\rm BG})\phi'/\mpl$ with $\phi_{\rm BG}$ being the background (asymptotic) value of the scalar. If the star's host galaxy is self-screened then this is the field value that minimizes the effective potential at mean galactic density, if the host galaxy is unscreened then the relevant density is the mean cosmic density. In theories that exhibit Vainshtein breaking the corresponding HSEE is \cite{Koyama:2015oma,Sakstein:2015aqx,Saito:2015fza,Jain:2015edg}
\begin{equation}\label{eq:HSEVAINBREAK}
\frac{\dd P(r)}{\dd r} = -\frac{G M(r)\rho(r)}{r^2}-\frac{\Upsilon_1 G\rho(r)}{4}\frac{\dd^2 M(r)}{\dd r^2},
\end{equation}
which can be expressed in alternate forms by taking derivatives of \eqref{eq:contstellar} to find
\begin{align}\label{eq:d2M}
\frac{\dd^2 M(r)}{\dd r^2}=8\pi r \rho(r) + 4\pi r^2\frac{\dd\rho(r)}{\dd r}.
\end{align}
The Vainshtein radius is necessarily several orders of magnitude larger than the radius of typical stars and so we do not give the HSEE for theories that do not include Vainshtein breaking.

These equations presented thus far do not form a closed set because the equation of state $P(\rho)$ is not known. One must either couple these equations to microphysical and macrophysical processes such as radiative transfer, nuclear burning, opacity, and convection to calculate the equation of state (EOS), or provide a known (or approximate) equation of state. Two important equations that will arise at various points in this section are the equation of radiative transfer, which describes the temperature gradient of the star due to photon transport:
\begin{equation}\label{eq:radtranseq}
\frac{\dd T}{\dd r} = -\frac{3}{4a}\frac{\kappa}{T^3}\frac{\rho L}{4\pi r^2},
\end{equation}
where $\kappa$ is the opacity, and the energy generation equation
\begin{equation}\label{eq:engen}
\frac{\dd L}{\dd r}=4\pi r ^2\sum_i \rho(r) \varepsilon_i(T,\rho).
\end{equation}
This equation describes the photon luminosity gradient produced by the interaction process $i$ with rate $\varepsilon_i$ per unit mass.

\subsubsection{Polytropic Models}

One simple and well-studied equation of state is the polytropic equation of state\cite{Chandrasekhar1939}
\begin{equation}\label{eq:polyeos}
P=K\rho^{\frac{n+1}{n}},
\end{equation}
which are good approximations for many stars, or at least some region of them. In the context of modified gravity (MG), polytropic equations of state allow one to decouple to gravitational and non-gravitational physics. This means one can discern the effects of changing the theory parameters without the need to account for possible degeneracies with non-gravitational processes. The stellar structure equations are self-similar for polytropic equations of state, which means one can work with dimensionless variables to extract the structure of the star independently of the central conditions. In particular, it is useful to work with the dimensionless radial coordinate
\begin{equation}\label{eq:rcdef}
r=r\ccc y,\,\textrm{ where } r\ccc^2\equiv\frac{(n+1)P\ccc}{4\pi G \rho\ccc^2},
\end{equation}
and $P\ccc$ and $\rho\ccc$ are the central pressure and density respectively. One can define the dimensionless function $\theta(y)$ via
\begin{equation}\label{eq:LEvars}
\rho=\rho\ccc\theta(y)^n\textrm{ and } P=P\ccc\theta(y)^{n+1}=K\rho\ccc^{\frac{n+1}{n}}\theta^n,
\end{equation}
which encodes the structure of the star. In GR, one can take a derivative of equation \eqref{eq:HSEGR} and apply equation \eqref{eq:contstellar} to find the Lane-Emden equation (LEE)
\begin{equation}\label{eq:LEGR}
\frac{1}{y^2}\frac{\dd}{\dd y}\left[y^2\frac{\dd \theta(y)}{\dd y}\right]=-\theta(y)^n.
\end{equation}
The equivalent equation for thin-shell screening theories is \cite{Davis:2011qf,Sakstein:2013pda,Sakstein:2015oqa}
\begin{equation}\label{eq:LECHAM}
\frac{1}{y^2}\frac{\dd}{\dd y}\left[y^2\frac{\dd \theta(y)}{\dd y}\right]=-\theta(y)^n
\begin{cases} 
      1 & y\le y_{\rm s} \\
      (1+2\beta^2(\phi_{\rm BG})) & y> y_{\rm s},
   \end{cases}
\end{equation}
where $\rs=r\ccc y_{\rm s}$ (i.e. $y_{s}$ is the dimensionless radius where screening begins) and the factor of $(1+2\beta^2(\phi_{\rm BG}))$ assumes that the star is fully unscreened outside the screening radius\footnote{If one were to attempt to go beyond this approximation and include the thin-shell factor $(1-M(\rs)/M(r))$ the self-similarity would be lost and, with it, the simplicity of the LEE. }. In Vainshtein breaking theories the LEE is \cite{Koyama:2015oma,Saito:2015fza}
\begin{equation}\label{eq:LEVAIN}
\frac{1}{y^2}\frac{\dd}{\dd y}\left[\left(1+\frac{n\Upsilon_1}{4}y^2\theta(y)^{n-1}\right)y^2\frac{\dd\theta }{\dd y}+\frac{\Upsilon_1}{2} y^2\theta(y)^n\right]=-\theta(y)^n,
\end{equation}
which has been derived using equation \eqref{eq:d2M} and using the relations \eqref{eq:LEvars}. The boundary conditions for the LEE are $\theta(0)=1$ ($P(r=0)=P\ccc$) and $\theta'(0)=0$ ($\dd P(r)/\dd r=0$ at the origin, which is a consequence of spherical symmetry). (See \cite{Horedt1987} for a detailed study of the LEE in GR.) One can find analytic solutions for specific values of $n$ but these are typically not relevant for astrophysics and so one must solve the LEE numerically. 

The radius of the star is defined as the radial coordinate where the pressure falls to zero, which defines $y_R$ such that $\theta(y_R)=0$. One then has
\begin{equation}\label{eq:RLE}
R=r\ccc y_R.
\end{equation}
The stellar mass can be found by integrating equation \eqref{eq:contstellar} to find
\begin{equation}\label{eq:massLE}
M=4\pi r\ccc^3\int_0^{y_R} y^2\theta(y)^n\dd y =\begin{cases} 
      \omega_R & \textrm{GR and Vainshtein breaking} \\
      \frac{\omega_R+2\beta^2(\phi_{
     \rm BG})\omega_s}{1+2\beta^2(\phi_{
     \rm BG})} & \textrm{Thin-shell}
   \end{cases},
\end{equation}
where we have replaced $\theta(y)^n$ using the appropriate Lane-Emden equation and defined
\begin{equation}
\omega_Y=\left.-y^2\frac{\dd\theta(y)}{\dd y}\right\vert_{y=Y}
\end{equation}
with $Y={\rm s}$ being short for $Y=y_{\rm s}$. 

Two important properties of polytopes that will be useful later one are the mass-radius relation
\begin{equation}\label{eq:MRpoly}
R=\gamma\left(\frac{K}{G}\right)^{\frac{n}{3-n}}M^{\frac{n-1}{n-3}};\quad\gamma\equiv(4\pi)^{\frac{1}{n-3}}\left(n+1\right)^{\frac{3}{3-n}}\omega_R^{\frac{n-1}{3-n}}y_R
\end{equation}
and the central density in terms of the mass and radius
\begin{equation}\label{eq:rhocpoly}
\rho_c=\delta\left(\frac{3M}{4\pi R^3}\right);\quad \delta=-\frac{y_R}{3\dd\theta/\dd y|_{y=y_R}}.
\end{equation}
These relations are derived in \cite{Chandrasekhar1939,Horedt2004} (and other similar textbooks). They apply to GR and Vainshtein breaking theories but not chameleon theories. We do not give the chameleon equivalents here since they will not be necessary\footnote{In fact, they have never been formally derived in the literature, although such a derivation is simple and straight forward.}. 

\subsubsection{Numerical Models: MESA}
\label{sec:MESA}

In order to model more complicated stars that do not have simple polytropic equations of state, one needs sophisticated numerical codes. One publicly available code that has proven invaluable for stellar structure in MG is MESA \cite{Paxton:2010ji}. MESA  solves the stellar structure equations coupled to the equations describing micro and macrophysical processes. 
The reader is referred to the instrumentation papers \cite{Paxton:2010ji,Paxton:2013pj,Paxton:2015jva} for a comprehensive review of MESA's capabilities. 

In the context of MG, MESA has been modified to solve the modified HSEE for both thin-shell screening (equation \eqref{eq:HSECHAM}) \cite{Chang:2010xh,Davis:2011qf,Sakstein:2013pda,Sakstein:2015oqa} and Vainshtein breaking theories (equation \eqref{eq:HSEVAINBREAK}) \cite{Koyama:2015oma}. MESA is a one-dimensional code (meaning that is assumes spherical symmetry) that splits each star into cells of varying lengths (the number of cells depends on the complexity of the star) and assigns relevant quantities (radius, density, temperature etc.) to each cell. The set of cells and these quantities then defines a stellar model at a specific time-step. Given a specific stellar model, the stellar structure equations are discretized on each cell solved to produce a new stellar model at a later time. Thus, the star is simulated over its entire lifetime. The publicly available version of MESA solves the GR HSEE \eqref{eq:HSEGR}. The modified versions of MESA solve either equation \eqref{eq:HSECHAM} or \eqref{eq:HSEVAINBREAK}. We will briefly describe how these modifications work below.

{\bf Thin-shell:} There are two independent chameleon modifications of MESA (see \cite{Najafi:2018apy} for a recent third). The first \cite{Chang:2010xh} solves the full scalar differential equation using a Gauss-Seidel relaxation algorithm. The second \cite{Davis:2011qf,Sakstein:2013pda,Sakstein:2015oqa}, uses the thin-shell approximation. Both codes agree very well but here we will only describe the latter implementation since it is more commonly used in the literature. Given a starting stellar model, the screening radius is computed by solving \cite{Chang:2010xh,Davis:2011qf,Sakstein:2013pda,Sakstein:2015oqa}
\begin{equation}\label{eq:findrs}
\chi_{\rm BG}\equiv \frac{\phi_{\rm BG}}{2\beta(\phi_{\rm BG})\mpl}=4\pi G\int_{\rs}^Rr\rho(r)\dd r.
\end{equation}
The code numerically integrates $r\rho(r)$ from the first cell until the cell where equation \eqref{eq:findrs} is satisfied. If the central cell is reached before this happens the screening radius is set to zero. In the latter case, the code simply rescales $G\rightarrow G(1+2\beta^2(\phi_{\rm BG}))$. In the former case, the mass inside the screening radius is found and used as an input for equation \eqref{eq:HSECHAM}. The next stellar model is then found by solving equation \eqref{eq:HSECHAM}. The screening radius is recomputed at every time-step to account for the changes in the star's structure. 

{\bf Vainshtein breaking:} MESA was first updated to include Vainshtein breaking by \cite{Koyama:2015oma,Sakstein:2015aqx}. In this case, the default (GR) HSEE is replaced by equation \eqref{eq:HSEVAINBREAK}. A numerical derivative of the density is taken by differencing across adjacent cells so that $\dd^2M(r)/\dd r^2$ can be computed in each cell using equation \eqref{eq:d2M}. The code then evolves to the next time-step using the modified HSEE for any input value of $\uo$, allowing the stellar evolution to be computed.  

\subsection{Radial Perturbations}

Moving away from equilibrium, one can consider Lagrangian perturbations so that
\begin{equation}\label{eq:stelpert1}
\vec{r}=r \hat{\vec{r}}+\vec{\delta r}
\end{equation}
and the velocity is $\vec{v}=\vec{\dot{\delta r}} $. The dynamics of $\vec{\delta r}$ describe perturbations of the star about its equilibrium configuration and, specializing to linear time-dependent radial perturbations\footnote{Non-radial modes are not important for thin-shell screened theories because they cannot be observed in galaxies other than our own (which is screened) and their governing equations have yet to be derived in MG theories.}
\begin{equation}
\delta r = |\vec{\delta r}| = \frac{\xi(r)}{r}e^{i\omega t}
\end{equation}
one can linearize the other quantities (pressure, density, etc.) and combine their governing equations to find, assuming GR for now,
\begin{equation}\label{eq:LAWE}
\frac{\dd}{\dd r}\left[r^4\Gamma_{1,0} P_0(r)\frac{\dd\xi(r)}{\dd r}\right] + r^3\frac{\dd}{\dd r}\left[\left(3\Gamma_{1,0}-4\right)P_0(r)\right]\xi(r) + r^4\rho_0(r)\omega^2\xi(r)=0,
\end{equation}
where subscript zeros refer to equilibrium quantities (found by solving the HSEE and other stellar structure equations) and $\Gamma_{1,0}=\dd\ln P_0/\dd\ln\rho_0$ is the first adiabatic index ($\Gamma_{1,0}=(n+1)/n$ for polytropic equations of state). Equation \eqref{eq:LAWE} is referred to as the \emph{linear adiabatic wave equation}. It is a Sturm-Liouville eigenvalue problem that must be solved given certain boundary conditions \cite{Cox1980} 
defined at the center 
and surface of the star 
The eigenfrequencies $\omega_n$ give the period of oscillation about the minimum $\Pi_n=2\pi/\omega_n$. Just like the equilibrium equations, the LAWE is self-similar and one can scale all of the dimensionful quantities out of the equation to find a dimensionless form in terms of a dimensionless frequency $\Omega^2 = \omega^2 R^3/(GM)$ so that the frequencies scale as
\begin{equation}
\omega_n^2\sim \Omega_n\frac{GM}{R^3} 
\end{equation}
or $\Pi\propto G^{-1/2}$. Theories where gravity is stronger therefore make stars of fixed mass and composition pulsate faster (or with a shorter period).

Reference \cite{Sakstein:2013pda} has derived the equivalent wave equation for thin-shell screened theories
\begin{align}
\label{eq:MLAWE}
\frac{\dd}{\dd r}\left[r^4\Gamma_{1,0} P_0(r)\frac{\dd\xi(r)}{\dd r}\right]& + r^3\frac{\dd}{\dd r}\left[\left(3\Gamma_{1,0}-4\right)P_0(r)\right]\xi(r)\nonumber\\& + r^4\rho_0(r)\left[\omega^2-8\pi\beta^2(\phi_{\rm BG})\rho_0(r)\Theta(r-\rs) \right]\xi(r)=0,
\end{align}
which is typically referred to as the \emph{modified linear adiabatic wave equation} (MLAWE). The boundary conditions are the same as in GR. One can see that the effect of the scalar field is to add a density-dependent mass term for $\xi(r)$ that increases $\omega$ (makes the period shorter) at fixed mass and composition, in line with our scaling arguments above. This is borne out by numerical simulations of polytropic and MESA models \cite{Sakstein:2013pda}. Another possible effect of stellar oscillations is that they may source scalar radiation, although detailed work for both non-relativistic\cite{Silvestri:2011ch,Upadhye:2013nfa} and relativistic stars \cite{Brax:2013uh} have found this to be negligible. 

For Vainshtein breaking theories, the derivation of the MLAWE is incredibly complicated but follows the relativistic derivation of \cite{Chandrasekhar1964} starting from perturbations of a relativistic gas sphere in a de Sitter background and taking the weak-field sub-horizon limit. The result is \cite{Sakstein:2016lyj}
\begin{align}
\label{eq:MLAWEVB}
&\frac{\dd}{\dd r}\left[r^4\left(\Gamma_{1,0} P_0(r)+\pi\uo G r^2\rho_0(r)^2\right)\frac{\dd\xi(r)}{\dd r}\right]+ r^3\frac{\dd}{\dd r}\left[\left(3\Gamma_{1,0}-4\right)P_0(r)\right]\xi(r)\nonumber\\& + r^4\rho_0(r)\omega^2\left(1-\frac{\pi\uo r^3\rho_0(r) M(r)}{2[M(r)+\pi r^3\rho_0(r)]^2} \right)\xi(r)=0,
\end{align}
with modified boundary condition at the center (see \cite{Sakstein:2016lyj}).

\subsubsection{Stellar Stability}

In GR, and thin-shell and Vainshtein breaking theories, the wave equation is a Sturm-Liouville eigenvalue equation of the form of a differential operator $\hat{\mathcal{L}}$ acting on a function $\xi(r)$ with weight function $W(r)$ i.e. $\hat{\mathcal{L}}\xi=W(r)\omega^2\xi$. This means we can bound the lowest eigenfrequency using the variational method by constructing the functional
\begin{equation}
\omega_0^2<F[\chi]\equiv\frac{\int_0^R\chi^*(r)\hat{\mathcal{L}}\chi(r)\dd r}{\int_0^R W(r)\chi^*(r)\chi(r)\dd r}
\end{equation}
for some trial function $\chi$. Taking this to be constant we find
\begin{equation}\label{eq:stabGR}
\omega_0^2<\frac{\int_0^R3r^2(3\Gamma_{1,0}-4)P_0(r)\dd r }{\int_0^R r^4\rho_0(r) }
\end{equation}
using the GR wave equation. When $\Gamma_{1,0}<4/3$ the lowest frequency is necessarily complex, signaling a tachyonic instability. In thin-shell screening theories, the equivalent of \eqref{eq:stabGR} is \cite{Sakstein:2013pda} 
\begin{equation}\label{eq:stabcham}
\omega_0^2<\frac{\int_0^R3r^2(3\Gamma_{1,0}-4)P_0(r)\dd r +\int_{\rs}^R 8\pi\beta^2(\phi_{\rm BG})Gr^4\rho_0(r)\dd r}{\int_0^R r^4\rho_0(r) }
\end{equation}
so that the instability is mitigated in a screening-dependent manner. Objects that are more unscreened can have $\Gamma_{1,0}<4/3$ and still be stable due to the compensating effect of the (positive) new term. This is borne out in the numerical computations of \cite{Sakstein:2013pda}. Finally, in Vainshtein breaking theories the expression is \cite{Sakstein:2016lyj}
\begin{equation}\label{eq:stabVB}
\omega_0^2<\frac{\int_0^R3r^2(3\Gamma_{1,0}-4)P_0(r)\dd r }{\int_0^R r^4\rho_0(r)\left(1-\frac{\pi\uo r^3\rho_0(r) M(r)}{2[M(r)+\pi r^3\rho_0(r)]^2} \right)}.
\end{equation}
When $\uo<0$ the instability is the same as in GR but when $\uo>0$ there is a second potential instability. For a star of constant density this always occurs when $\uo>49/6$. For more general models, one needs to integrate over the equilibrium structure to determine the presence of the instability, although, given the large value for constant density stars, it is unlikely that the instability is realized in practice for sensible choices of $\uo$. 

\section{Stellar Structure Tests}

\label{sec:SStests}

In this section we review different objects that the theory developed in the last section has been applied to and the resulting bounds on screened MG theories.

\subsection{Main-Sequence Stars}

\label{sec:MS}
\subsubsection{The Eddington Standard Model}

One of the simplest treatment of main-sequence stars which works well for low-mass objects is the \emph{Eddington standard model}, which makes the assumption that the star is supported by a combination of radiation pressure from photons generated by nuclear burning in the core and hydrodynamic gas pressure (ideal gas law):
\begin{equation}\label{eq:Prad}
P_{\rm rad}=\frac{1}{3}a T^4\textrm{  and  } P_{\rm gas}= \frac{k_{\rm B}\rho T}{\mu m_{\rm H}},
\end{equation}
where $m_{\rm H}$ is the mass of a hydrogen atom and $\mu$ is the mean molecular weight (number of particles per atomic unit). Introducing $b={P_{\rm gas}}/{P}$, equation \eqref{eq:Prad} implies that
\begin{equation}\label{eq:ESMbeq}
\frac{b}{1-b}=\frac{3k_{\rm B}\rho}{a\mu m_{\rm H}T^3}.
\end{equation}
This implies $b$ is a constant if one makes the approximation that the specific entropy ($s\propto \rho/T^3$) is constant. The total pressure is then
\begin{equation}
P=K(b)\rho^{\frac{4}{3}}\textrm{  with  } K(b)=\left(\frac{3}{a}\right)^{\frac{1}{3}}\left(\frac{k_{\rm B}}{\mu m_{\rm H}}\right)^{\frac{4}{3}}\left(\frac{1-b}{b^4}\right)^{\frac{1}{3}}
\end{equation}
so that the star is therefore polytropic with $n=3$ and its structure can be found by solving the Lane-Emden equation for the theory of gravity in question.


For MG, the most important quantity for main-sequence stars is the luminosity, which must be determined from the radiative transfer equation. (In this section we assume that the opacity is constant, which is a good approximation for main-sequence stars where the dominant contribution comes from electron scattering.) Differentiating equation \eqref{eq:Prad}, one can find an expression for the surface luminosity using the appropriate HSEE (\eqref{eq:HSEGR} for GR, \eqref{eq:HSECHAM} for thin-shell models, and \eqref{eq:HSEVAINBREAK} for Vainshtein breaking) \begin{equation}\label{eq:luminosityESM}
L=\frac{4\pi G M(1-b)}{\kappa}\begin{cases} 
      1 & \textrm{GR and Vainshtein breaking} \\
     1+2\beta^2(\phi_{\rm BG}) \left(1-\frac{M(\rs)}{M}\right)& \textrm{thin-shell},
   \end{cases},
\end{equation}
where $M=M(R)$ is the stellar mass. Thus, in order to determine the luminosity (at fixed mass) we must calculate $b$. This is accomplished by inserting the definition of $r\ccc$ (equation \eqref{eq:rcdef}) into equation \eqref{eq:massLE} to find a quartic equation (Eddington's quartic equation) \cite{Davis:2011qf,Koyama:2015oma}:
\begin{equation}\label{eq:eddingtonquartic}
\frac{1-b}{b^4}=\left(\frac{M}{M_{\rm Edd}}\right)^2\begin{cases} 
      1 & \textrm{GR} \\
      \left(\frac{\bar{\omega}_R}{\omega_R}\right)^2 & \textrm{Vainshtein Breaking}\\
     \left[(1+2\beta^2(\phi_{\rm BG}))\frac{\bar{\omega}_R}{\omega_R+2\beta^2(\phi_{\rm BG})\omega_{\rm s}}\right]^{\frac{2}{3}}& \textrm{Thin-shell}
   \end{cases},
\end{equation}
where $\bar{\omega}\approx2.018$ is the GR value and the Eddington mass is
\begin{equation}\label{eq:Medd}
M_{\rm Edd}=\frac{4\bar{\omega}_R}{\sqrt{\pi}G^{\frac{3}{2}}}\left(\frac{k_{\rm B}}{\mu m_{\rm H}}\right)^2\left(\frac{3}{a}\right)^{\frac{1}{2}}\approx 18.3 M_\odot \mu^{-2}.
\end{equation}
Note that the GR and Vainshtein breaking luminosities are not identical despite having the same expression since $b$ is determined from different equations. 

At this point, one can discern the gross effects of MG on the stellar luminosity. First, note from equation \eqref{eq:luminosityESM} that when $b=1$ the luminosity is zero. This is because this extreme value corresponds to no radiation pressure and hence no photons. When $b\ll 1$ the star is dominated by radiation pressure and one has $L\propto GM$. Conversely, when $b$ is close to unity (so that the star is gas-pressure supported) one can write $b=1-\delta$ for $\delta\ll1$ and equation \eqref{eq:eddingtonquartic} shows that $\delta\propto(M/M_{\rm Edd})^2\propto G^4 M^2$. One then has $L\propto G^4M^3$. This means that the effects of MG are more pronounced in pressure supported stars. Equation \eqref{eq:eddingtonquartic} requires $b\approx 1$ for $M<M_{\rm Edd}$ whereas $b\ll1$ for $M>M_{\rm Edd}$ so that low-mass stars are gas-supported and high-mass stars are pressure supported. We therefore expect the effects of MG to be more pronounced in low-mass stars.

The procedure for calculating the luminosity in any given gravity theory is as follows: first, one numerically solves the relevant $n=3$ Lane-Emden equation for a given set of parameters (there are no free parameters in GR) to find $\omega_{R}$ ($=\bar{\omega}_{R}$ in GR). For thin-shell models, one must also find the screening radius and $\omega_{\rm s}$ using \eqref{eq:findrs} (see \cite{Davis:2011qf,Sakstein:2015oqa} for the details). 
Once $\omega_R$ (and $\omega_{\rm s}$ for chameleons) have been obtained, equation \eqref{eq:eddingtonquartic} can be solved numerically to find $b$. This can then be put into \eqref{eq:luminosityESM} to find the luminosity.

\begin{figure}
\centering
{\includegraphics[width=0.45\textwidth]{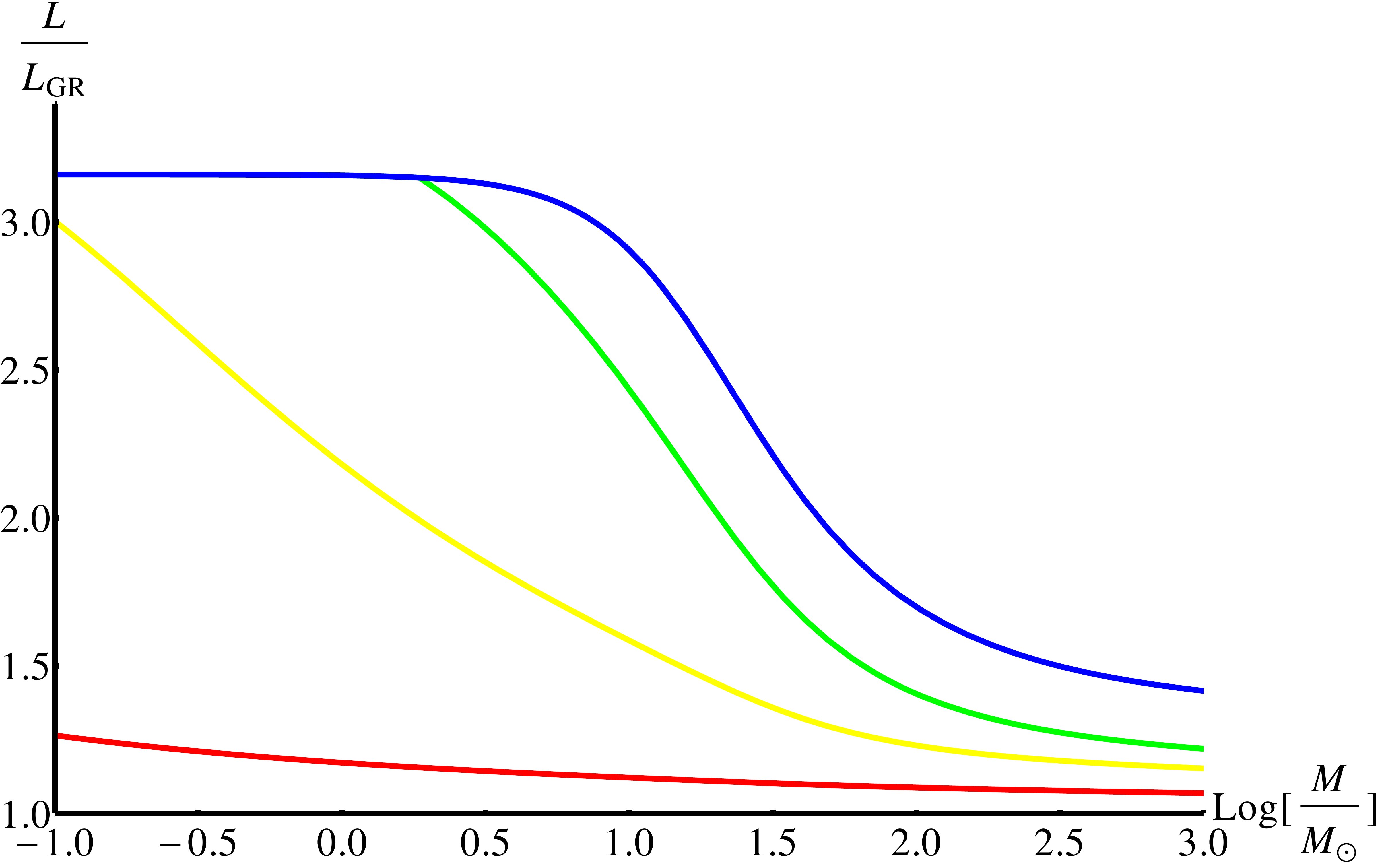}}
{\includegraphics[width=0.45\textwidth]{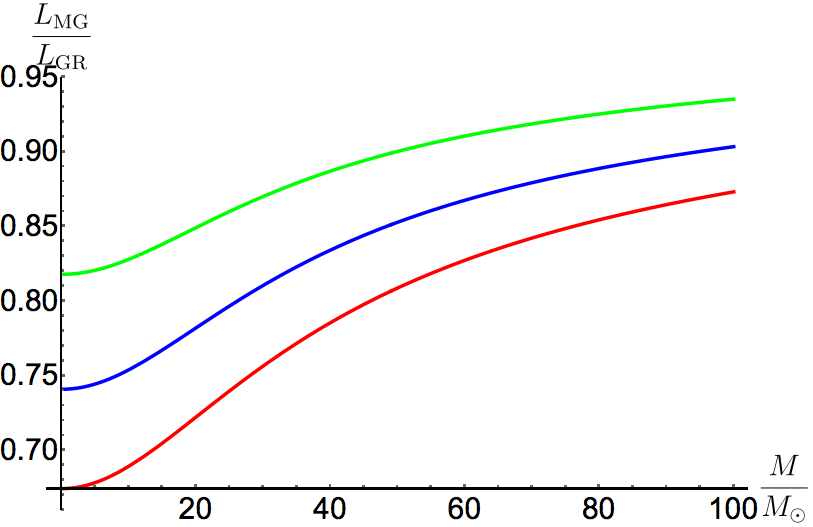}}
\caption{The luminosity enhancement for main-sequence stars assuming the Eddington standard model. \emph{Left panel}: Thin-shell screening theories (taken from \cite{Davis:2011qf}). The plot shows the enhancement for $\beta(\phi_{\rm BG})=1/\sqrt{6}$ (corresponding to $f(R)$ theories) with (from top to bottom) $\chi_{\rm BG} = 10^{-4}$ (blue), $10^{-5}$ (green), $5\times10^{-6}$ (yellow), and $10^{-6}$ (red). \emph{Right panel}: Vainshtein breaking theories (taken from \cite{Koyama:2015oma}). From top to bottom, $\Upsilon_1=0.4$ (green), $\Upsilon_1=0.6$ (blue), and $\Upsilon_1=0.8$ (red).}\label{fig:ESMplot}
\end{figure}

Plots of the ratio $L/L_{\rm GR}$ are shown in figure \ref{fig:ESMplot} for both thin-shell and Vainshtein breaking theories. In both cases $\mu=1/2$, appropriate for hydrogen stars. Evidently, the effects of MG are indeed more pronounced in low-mass objects due to their gas pressure support. We have chosen $\beta(\phi_{\rm BG})=1/\sqrt{6}$ for thin-shell models corresponding to the (constant) value predicted by $f(R)$ models. When $\chi_{\rm BG}\ge10^{-5}$ the enhancements plateau at low masses because the stars are fully unscreened. The asymptotic value is precisely $(1+2\beta(\phi_{\rm BG})^2)^4=(4/3)^4\approx 3.16 $, in agreement with our prediction above that $L\propto G^4$ for full unscreened gas pressure supported stars. We chose $\Upsilon_1>0$ for the Vainshtein breaking models, which, evidently, lowers the luminosity compared with GR. A good rule of thumb (but by no means a concrete feature) is that positive values of $\Upsilon_1$ weaken gravity (compared with GR) in the Newtonian limit\footnote{This is because $\dd^2M/\dd r<0$ for low mass homogeneous stars.}. Had we chosen $\Upsilon_1<0$ we would have found the converse behavior i.e. the luminosity would have been enhanced, a consequence of strengthened gravity. 

\subsection{MESA Models}

\subsubsection{Thin-Shell Stars}

\begin{figure}
\centering
\includegraphics[width=0.6\textwidth]{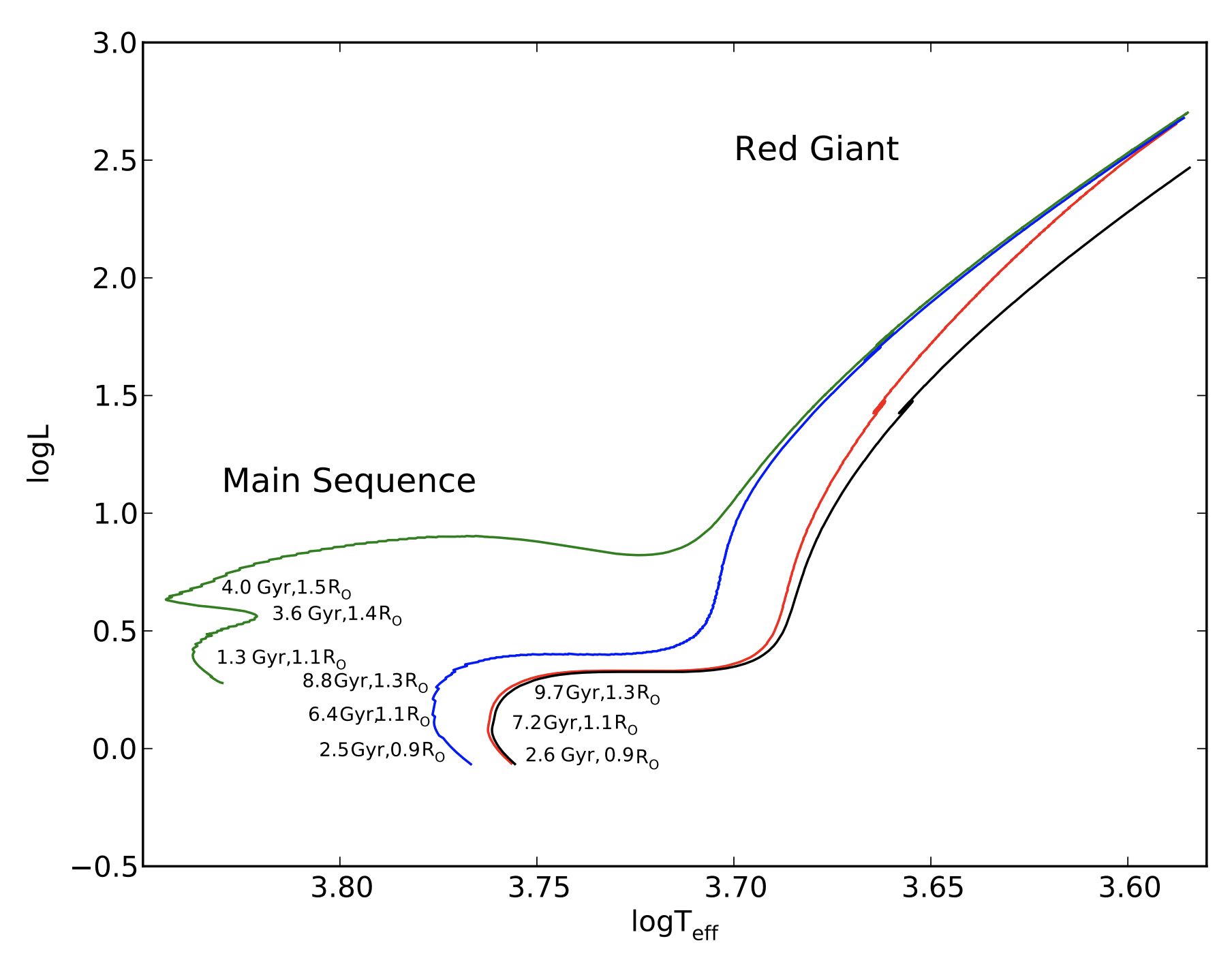}
\caption{The HR tracks of a solar mass star of solar metallicity ($Z=0.02$). From bottom to top: GR (black), $10^{-7}$ (red), $10^{-6}$ (blue), $5\times 10^{-6}$ (green). In each case, the value of $\beta(\phi_{\rm BG})=1/\sqrt{6}$ so that the chameleon theory can be re-written as an $f(R)$ theory. The radius and age at the points where the central hydrogen mass fraction has fallen to $0.5$, $0.1$ and $10^{-5}$ are indicated in the figure (from bottom to top) with the exception of the red curve, which mimics GR on the main-sequence. Figure taken from \cite{Davis:2011qf}.}\label{fig:cham_MESA}
\end{figure}

For thin-shell screening theories, references \cite{Chang:2010xh,Davis:2011qf} have used a modified version of MESA to compute the color-magnitude or Hertzprung-Russell (HR) tracks for solar mass stars. An example of this is shown in figure \ref{fig:cham_MESA}. The curves show the evolution of a solar mass and metallicity star from the zero-age main-sequence to the tip of the red giant branch in $f(R)$ chameleon theories ($2\beta^2(\phi_{\rm BG})=1/3$). Also shown are the radius and ages of the star when the central hydrogen mass fraction $X=0.5$, $0.1$, and $10^{-5}$ so that one can compare stars at similar points in their evolution. The parameters are chosen so that the stars are progressively more unscreened from bottom to top. The curve at $\chi_{\rm BG}=10^{-7}$ mimics GR on the main-sequence because the star is fully screened (recall main-sequence stars have $\pn\sim 10^{-6}$) but becomes unscreened on the red giant branch when the radius of the star increases about a factor of $10$, lowering its Newtonian potential. The blue curve has a comparable shape to GR but is shifted to higher temperatures and luminosities, indicating that the star is brighter and hotter than its GR counterpart. The green curve corresponds to a star that is fully unscreened, and looks like the HR track for a $2M_\odot$ star. In all cases, at fixed $X$ more unscreened stars are younger, indicating that stellar evolution has proceeded at a faster rate. This is because the amount of nuclear fuel is fixed (at fixed mass) but more unscreened stars need to consume it at a faster rate in order to combat the increased gravity. Thin-shell screening stars are therefore hotter, brighter, and more ephemeral the more unscreened they are. Unfortunately, these predictions have yet to be utilized as a test of chameleon theories. The main reason for this is that one requires unscreened galaxies---dwarf galaxies in cosmic voids---in order for the stars to become sufficiently unscreened. Main-sequence and post-main-sequence stars are typically not resolvable in such galaxies.

\subsubsection{Vainshtein Breaking Stars}

\begin{figure}
\centering
{\includegraphics[width=0.45\textwidth]{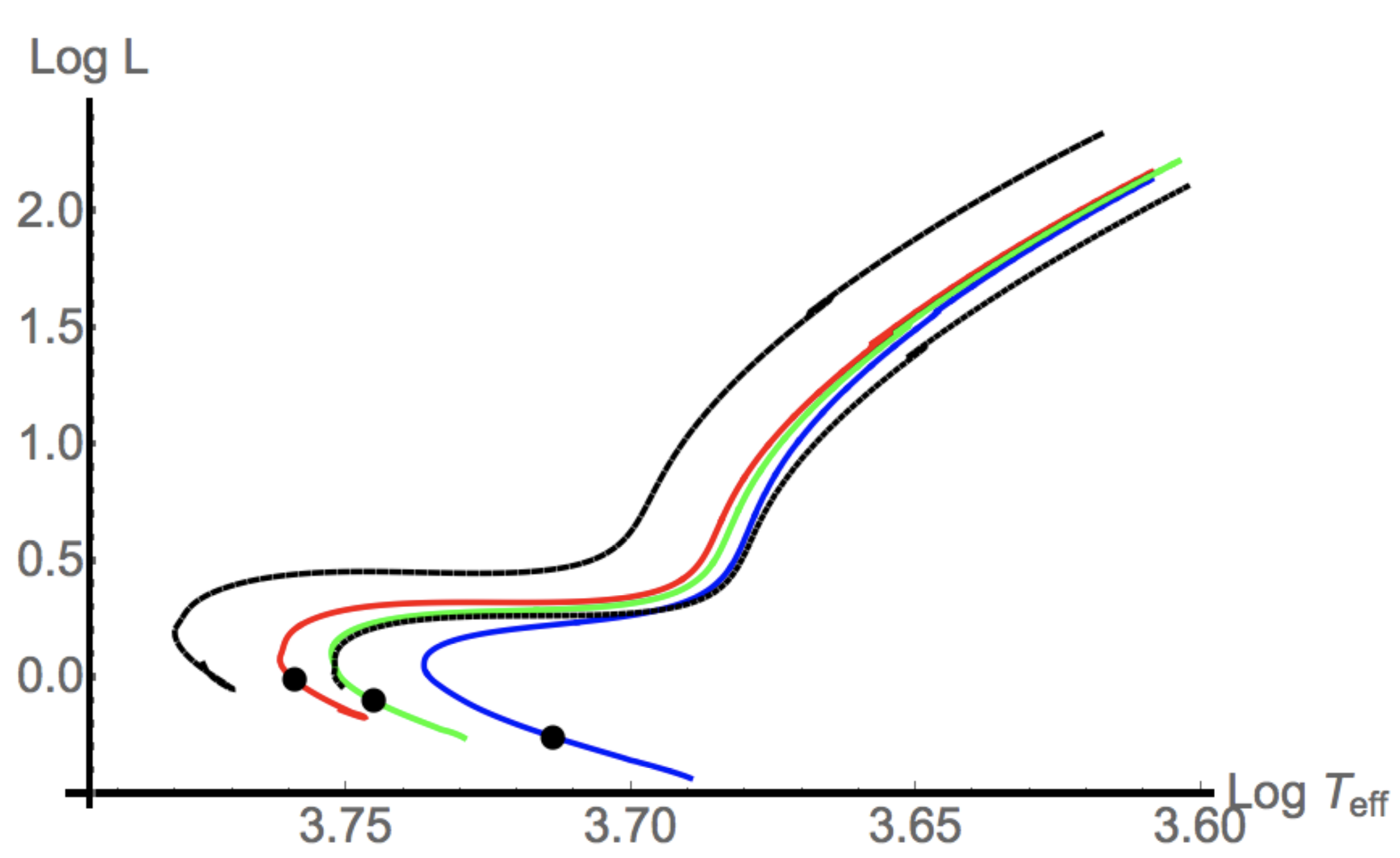}}
{\includegraphics[width=0.45\textwidth]{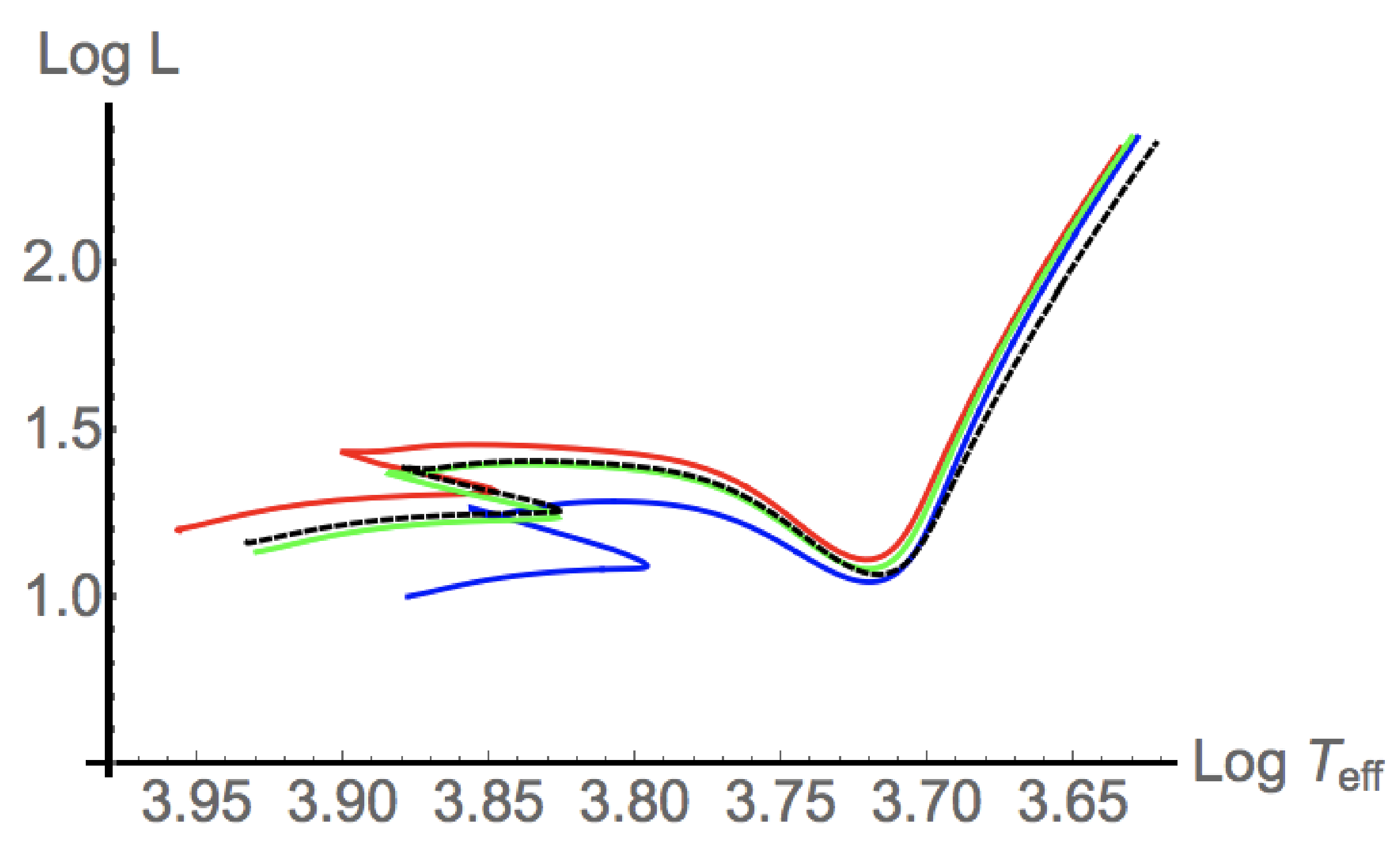}}
\caption{HR tracks in GR and Vainshtein breaking theories. From top to bottom: GR, $Z=0.02$ (red, solid), GR, $ Z=0.03$ (black, dotted), $\uo=0.1$, $Z=0.02$ (green, solid), $\uo=0.3$, $Z=0.02$ (blue, solid). \emph{Left Panel}: One solar mass. The solid circle shows the star when its age is $4.6\times 10^{9}$ yr. \emph{Right Panel}: Two solar masses. Figures taken from \cite{Koyama:2015oma}.}\label{fig:VB_MESA}
\end{figure}

The HSEE for Vainshtein breaking theories was implemented into MESA by \cite{Koyama:2015oma} using the method outlined in section \ref{sec:MESA}. The HR tracks for solar mass stars and two solar mass stars are shown in the left and right hand panels of figure \ref{fig:VB_MESA} respectively. One can see that, at fixed metallicity the effects of increasingly positive $\uo$ is to make the star dimmer and cooler. This is because positive values of $\uo$ act in an equivalent manner to weakening gravity and therefore the star needs to burn nuclear fuel at a slower rate to stave off gravitational collapse. Another consequence of this is that stars evolve more slowly when $\uo$ is more positive, as evidenced by the location of the filled circles in the left panel. Negative value of $\uo$ have the opposite effect (i.e. gravity is strengthened); fuel is consumed at a faster rate and the star is hotter, brighter, and more ephemeral. On the main-sequence, these effects are degenerate with metallicity; it is evident from the figures that a GR $Z=0.03$ star has a similar main-sequence track to a Vainshtein breaking star with $\uo=0.1$ and $Z=0.02$. (If $\uo<0$ the effects of Vainshtein breaking are degenerate with decreasing the metallicity.) This degeneracy vanishes on the red giant branch. In theory, the effects of Vainshtein breaking should be present in all stars in our local neighborhood. In practice, to date there have been no local tests, either proposed or performed. This is due partly to the degeneracy with metallicity, although this can either be corrected for with other measurements or avoided by using post-main-sequence stars.

\subsection{A Stellar Bound for Vainshtein Breaking Theories}

One important requirement for the stability of stars is that $P''(r)<0$ \cite{Delgaty:1998uy}. At the center of the star, the pressure, density, and mass can be expanded as
\begin{equation}
P(r)=P\ccc-\frac{P_2}{2}r^2+\cdots,\,\, \rho(r)=\rho\ccc-\frac{\rho_2}{2}r^2+\cdots,\,\textrm{ and } M(r)=\frac{M_3}{3}r^3+\cdots,
\end{equation}
where the linear terms are absent in the expansions of $P(r)$ and $\rho(r)$ because one needs $P'(0)=\rho'(0)=0$; the expansion for $M(r)$ begins at cubic order in order to be consistent with equation \eqref{eq:contstellar}. Plugging these expansions into the HSEE \eqref{eq:HSEVAINBREAK} one finds
\begin{equation}
P_2=\frac{G\rho_cM_3}{3}\left(1+\frac{3\uo}{2}\right)<0,
\end{equation}
implying the bound $\uo>-2/3$. This bound was first derived by \cite{Saito:2015fza} using similar arguments.

\subsection{Dwarf Stars}
\label{sec:DS}
Dwarf stars are those that populate the mass range between Jupiter mass planets ($M_{\rm J}\sim10^{-3}M_\odot$) and main-sequence stars with masses $M\sim\mathcal{O}(0.1 M_\odot)$. When first formed, a star will contract under its own self-gravity liberating energy and increasing the temperature and density. The contraction must be halted by the onset of pressure support either due to electron degeneracy pressure or thermonuclear fusion. In the former case, the star is inert and is referred to as a \emph{brown dwarf}. In the latter case, it is a \emph{red dwarf}. Only stars that are sufficiently heavy can achieve the requisite core density and temperature for hydrogen burning to proceed efficiently. Thus, low-mass stars are brown dwarfs and higher mass stars are red dwarfs. The transition mass, the minimum mass for hydrogen burning (MMHB), is $M_{\rm MMHB}\approx 0.08 M_\odot$ in GR. A detailed account of low-mass stars can be found in \cite{Burrows:1992fg}. In the context of MG, dwarf stars are good probes of Vainshtein breaking theories \cite{Sakstein:2015zoa,Sakstein:2015aac}, and so we focus exclusively on these in this subsection.

\begin{figure}
\centering
\includegraphics[width=0.5\textwidth]{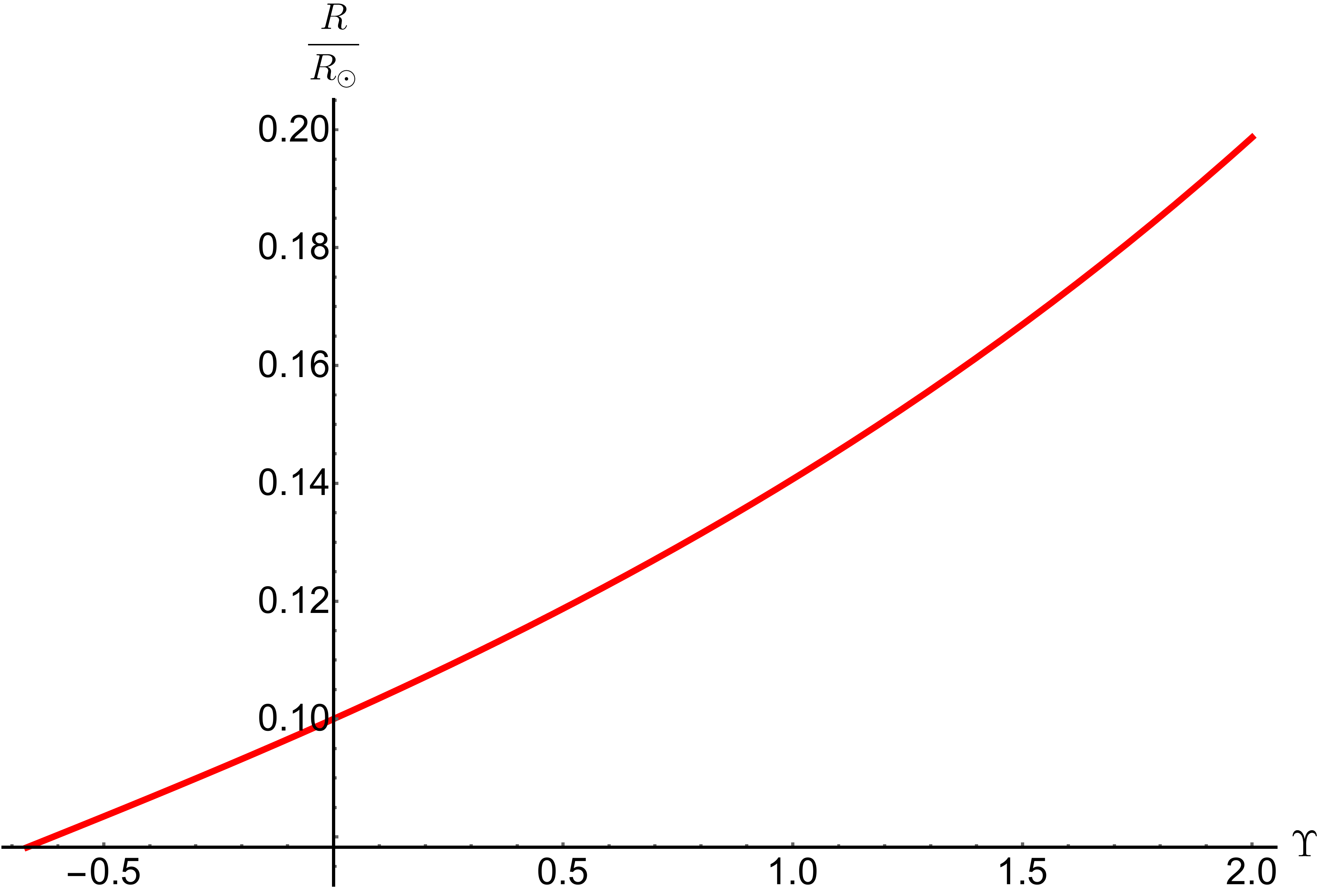}
\caption{The radius of brown dwarf stars in Vainshtein breaking theories as a function of $\Upsilon_1$. Taken from \cite{Sakstein:2015aac}.}\label{fig:BDradius}
\end{figure}

\subsubsection{Brown Dwarf Stars: The Radius Plateau}
\label{sec:BD}
Brown dwarfs are inert (non-hydrogen burning) stars\footnote{Higher mass brown dwarfs may burn deuterium or lithium (or both) for a short time (until the reserve is depleted). In fact there are minimum masses for deuterium and lithium burning analogous to the MMHB. } composed primarily of molecular hydrogen and helium in the liquid metallic phase with the exception of a thin layer near the surface, which is composed of a weakly coupled plasma that is well-described by the ideal gas law. They are fully convective and therefore contract along the Hyashi track with a polytropic $n=1.5$ EOS \cite{Kippenhahn1994}. In fact, coulomb corrections to the electron scattering processes shift the EOS of lower mass brown dwarfs ($M\lsim 4M_{\rm J}$) to lower values $n\approx 1$ \cite{Burrows:1992fg,Saumon1995}. For $n=1$ one has $P\ccc=K\rho\ccc^2$ (c.f. equation \eqref{eq:LEvars} and recall $\theta(0)=1$) so that equation \eqref{eq:rcdef} gives $r\ccc^2=K/(2\pi G)$ and the radius, $R=r\ccc y_R$ is, is independent of the mass. This leads to a \emph{radius plateau} in the mass-radius relation for stars with masses $M_{\rm J}<M<M_{\rm MMHB}$. In GR, the plateau lies at $R\approx0.1 R_\odot$ but in Vainshtein breaking theories $y_R$ depends on $\Upsilon_1$ and therefore so does the plateau radius. This is shown in figure \ref{fig:BDradius}. One can see that the changes in the radius are significant for $|\Upsilon_1|\sim\mathcal{O}(1)$, although whether this can be used to place new bounds is not clear since the data pertaining to the radius plateau is currently sparse \cite{Chabrier:2008bc}. Future data releases from Gaia, Kepler, or their successors may be able to populate the brown dwarf mass-radius diagram sufficiently.

\subsubsection{Red Dwarf Stars: The Minimum Mass for Hydrogen Burning}
\label{sec:RD}
The central conditions in low-mass stars are not sufficient for efficient burning on the PP-chains. In particular, the coulomb barrier for the $^{\rm 3}\textrm{He}$-$^{\rm 3}\textrm{He}$ and $^{\rm 3}\textrm{He}$-$^{\rm 4}\textrm{He}$ cannot be overcome at the relevant central temperatures and densities ($10^6$ K and $10^3$ g/cm$^3$). Instead, proton burning proceeds via deuterium burning with the end point being Helium-3. 
The MMHB is the smallest mass where the luminosity generated by this reaction process can balance the luminosity lost from the star's surface. 

A simple model of red dwarf stars first presented in \cite{Burrows:1992fg} for GR was adapted for Vainshtein breaking theories by \cite{Sakstein:2015zoa,Sakstein:2015aac} who showed that the MMHB is sensitive to $\Upsilon_1$. In this model, the star is supported by a combination of degeneracy pressure and the ideal gas law, which are both described by $n=1.5$ polytropic equations of state. Stable hydrogen burning is achieved when 
\begin{equation}\label{eq:MMHB}
3.76M_{-1}=\left[\frac{\left(1+\frac{\Upsilon_1}{2}\right)}{\kappa_{-2}}\right]^{0.11}\left(1+\frac{3\Upsilon_1}{2}\right)^{0.14}\frac{\gamma^{1.32}\omega_R^{0.09}}{\delta^{0.51}}I(\eta);\quad I(\eta)=\frac{(\alpha+\eta)^{1.509}}{\eta^{1.325}},
\end{equation}
where $M_{-1}=M/(0.1 M_\odot)$, $\kappa_{-2}=\kappa_{R}/10^{-2}$ with $\kappa_{\rm R}$ being the Rosseland mean opacity, $\alpha=4.82$, and $\gamma$ and $\delta$ are defined in \eqref{eq:MRpoly} and \eqref{eq:rhocpoly} respectively. The degeneracy parameter $\eta$ is the ratio of the Fermi energy to $k_{\rm B}T$ and measures the relative contribution of each type of pressure, degenracy pressure being more important for larger $\eta$. 

\begin{figure}
\centering
\includegraphics[width=0.6\textwidth]{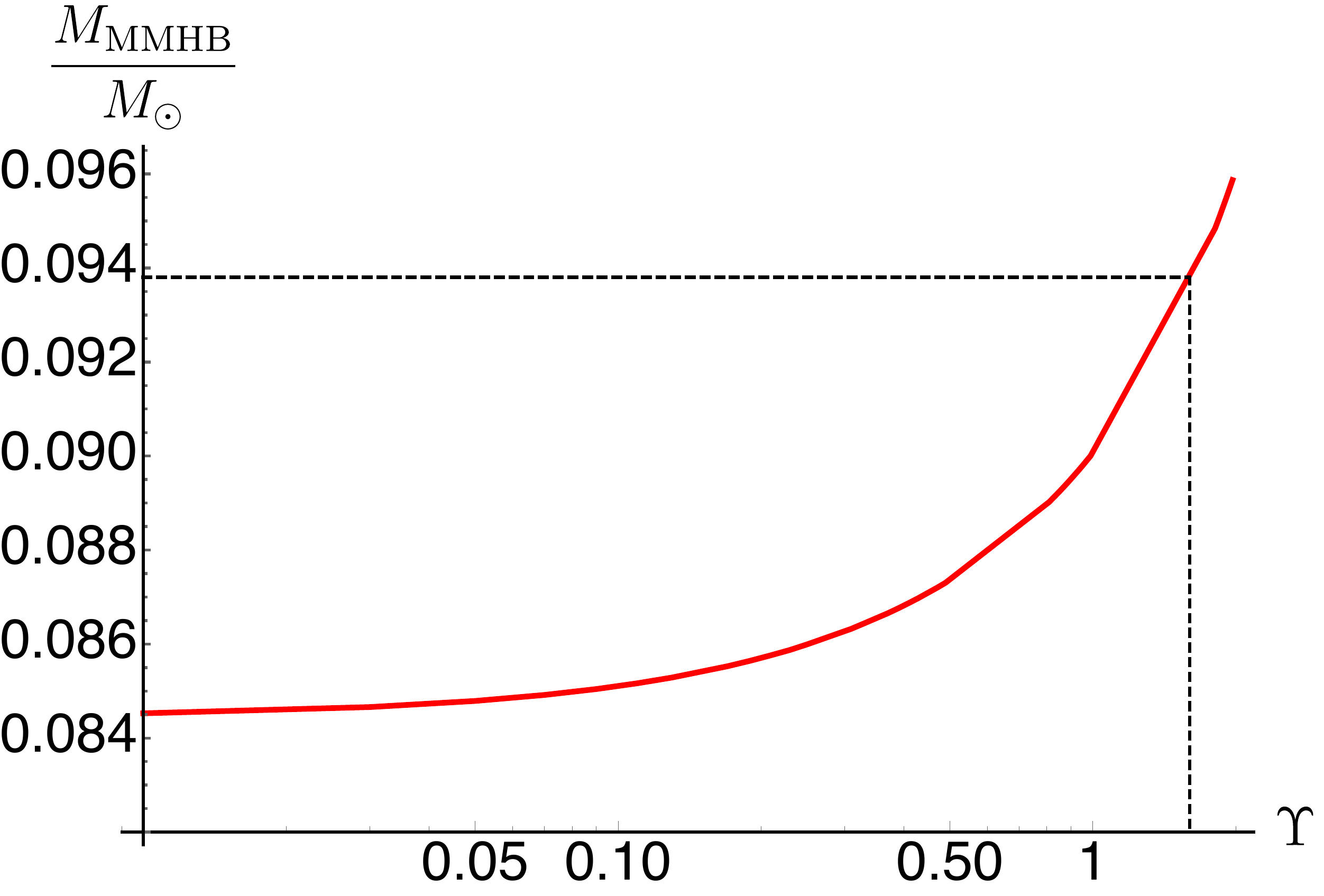}
\caption{The MMHB as a function of $\Upsilon_1$. The black dashed line shows the upper limit on the mass of the lightest red dwarf presently observed and the corresponding value of $\Upsilon_1$. Figure taken from \cite{Sakstein:2015zoa}.}
\label{fig:MMHB}
\end{figure}

The function $I(\eta)$ has a minimum value of $2.34$ at $\eta=34.7$ and so there is a minimum value of $M$ for which \eqref{eq:MMHB} can be satisfied, the MMHB. Assuming $\kappa_{-2}=1$ (we will discuss this later), the MMHB in GR is $M_{\rm MMHB}^{\rm GR}\approx 0.08M_\odot$ whereas in Vainshtein breaking theories it depends on $\Upsilon_1$ as shown in figure \ref{fig:MMHB}. One can see that, for positive values of $\Upsilon_1$, the MMHB is larger than the GR value. This is because the weakened gravity results in lower central densities and temperatures at fixed mass so that heavier objects are needed to reach the requisite conditions for hydrogen burning. One cannot take theories with $\Upsilon_1$ too large because the theory would predict that observed red dwarf stars should be brown dwarfs. Indeed, the lightest red dwarf (M-dwarf) is Gl 866 C with a mass $M=0.0930\pm0.0008_\odot$ \cite{Segransan:2000jq}. Vainshtein breaking theories are only compatible with this observation if the bound $\Upsilon_1<1.6$ is satisfied.

This bound is incredibly robust. Indeed, there are few degeneracies with other astrophysical effects. There is a degeneracy with the opacity but, as is evident in equation \eqref{eq:MMHB}, this is very mild and is not strong enough to impart any uncertainty onto this bound. Similarly, variations in the chemical composition between different dwarf stars are small and the compositions themselves do not evolve significantly over the life-time of the star. Another possible degeneracy is rotation, but this acts to increase the MMHB \cite{Salpeter1955,Kippenhahn1970} and can therefore only make the bound stronger. Finally, the method used to infer the star's mass is insensitive to the theory of gravity. The mass is either inferred from empirical relations, which do not assume any gravitational physics, or from the orbital dynamics of binaries \cite{Henry:1993tk}, which occurs in a regime where there is no Vainshtein breaking (i.e. outside the objects) so that the equations are identical to GR. See \cite{Sakstein:2015zoa,Sakstein:2015aac} for an extended discussion on this.

\subsection{White Dwarf Stars: the Chandrasekhar Mass and Mass-Radius Relation}
\label{sec:WD}
White dwarf stars are the remnants of low-mass stars ($M\lsim 8M_\odot$) that have gone off the main-sequence to become giant stars and have subsequently had their outer layers blown away by stellar winds leaving only the core. In the absence of any thermonuclear fusion, electron degeneracy pressure provides the counter-gravitational support. Low mass white dwarf stars are well described by $n=1.5$ polytropic equations of state ($P\propto \rho^{\frac{5}{3}}$) corresponding to a non-relativistic gas whereas high-mass white dwarfs are best described by $n=3$ ($P\propto\rho^{\frac{4}{3}}$) corresponding to a fully relativistic gas. Following equation \eqref{eq:MRpoly}, this means that low-mass white dwarfs follow the mass-radius relation $R\propto M^{-\frac{1}{3}}$ whereas fully relativistic white dwarfs have a fixed mass (the Chandrasekhar mass). If one tries to go to higher masses, the star is unstable and a thermonuclear explosion occurs, resulting in a type Ia supernova. This is the same instability found using perturbation theory (see equation \eqref{eq:stabGR}). 

The majority of white dwarf stars are composed primarily of $^{12}\textrm{C}$, for which an equation of state can easily be found. We will follow the method of \cite{Shapiro1983}, which \cite{Jain:2015edg} have adapted to Vainshtein breaking theories. Defining $x=p_{\rm F}/m_{\rm e}$ where $p_{\rm F}$ is the Fermi momentum, the number density of degenerate electrons is
\begin{equation}
n_{\rm e}=\frac{m_{\rm e}^3x^3}{3\pi^2}
\end{equation}
while the electron pressure and energy density are $P_{\rm e}=m_{\rm e}^4\Psi_1(x)$ and $\epsilon_{\rm e}=m_{\rm e}^4\Psi_2(x)$ with $\Psi_i(x)$ given in \cite{Shapiro1983}.
The density receives contributions from both the carbon atoms and the electrons but the former far heavier than the latter and so the density is $\rho\approx\rho_{\rm C} $.
On the other hand, the pressure comes primarily from the electrons and so one has $P\approx P_{\rm e}$. One can use these approximations with the appropriate HSEE to construct white dwarf models. In this case, the unknown functions are $m(r)$ and $x(r)$ which satisfy $M(0)=0$ and $x(0)=x_0$ so that their is one free parameter defined at the center for the star. The radius is defined by $P(R)=0$ (which implies $x=0$) so that $M=M(R)$. Varying $x_0$ allows one to build up the mass-radius relation.

\begin{figure}
\centering
{\includegraphics[width=0.32\textwidth]{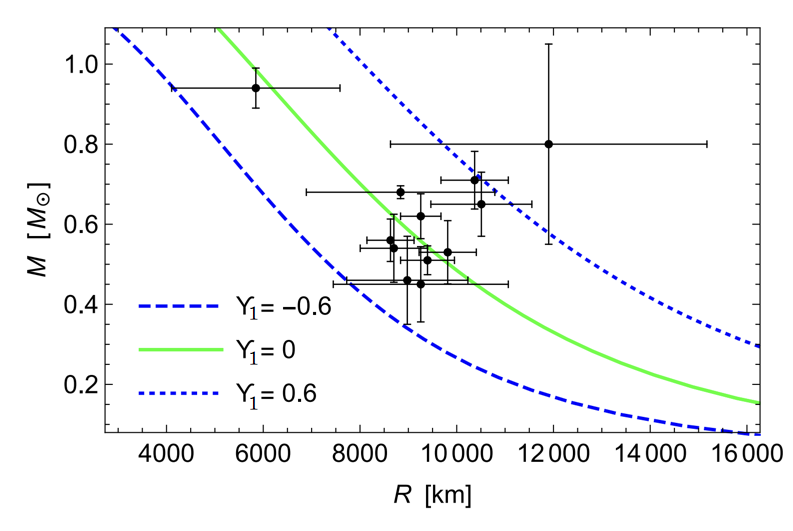}}
{\includegraphics[width=0.32\textwidth]{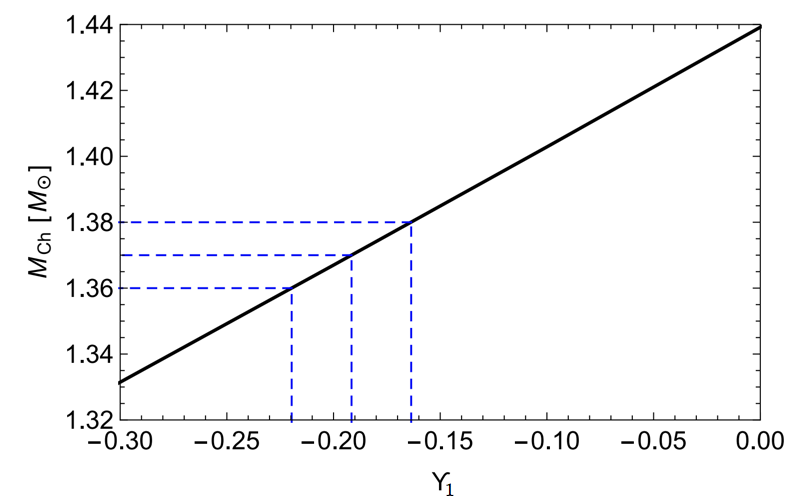}}
{\includegraphics[width=0.32\textwidth]{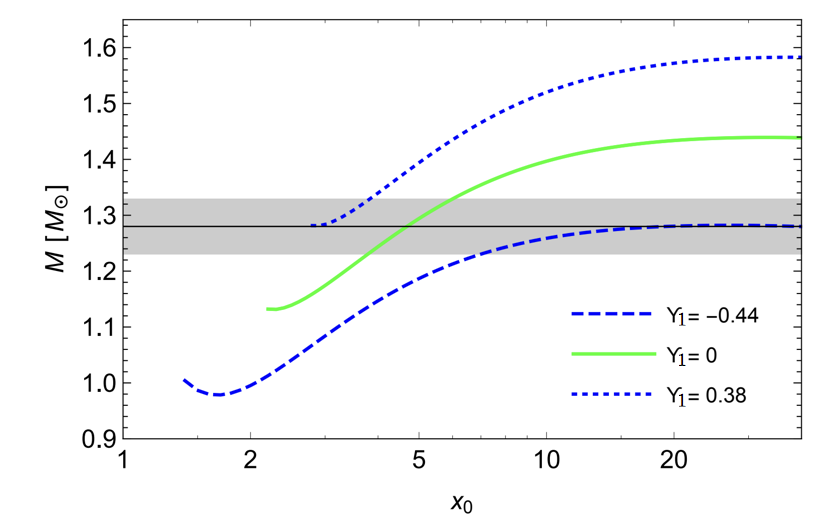}}
\caption{\emph{Left Panel}: The mass-radius relation for white dwarf stars in GR (green, center, solid) and Vainshtein breaking theories with $\Upsilon_1=-0.6$ (blue, dotted, upper) and $\Upsilon_1=0.6$ (blue, dashed, lower). Also shown are the masses and radii of 12 observed white dwarfs from \cite{Holberg:2012pu} that were used by \cite{Jain:2015edg} for their analysis. \emph{Central Panel}: The Chandrasekhar mass as a function of $\uo$ for $\uo<0$ (black, solid). The blue dashed lines show the values of $\uo$ that correspond to the central value and upper and lower limits of the mass of the heaviest observed white dwarf ($M=1.37\pm0.01M_\odot$). \emph{Right Panel}: The mass-central density (parameterized by $x_0$) relation for $GR$ (green, center, solid), $\uo=-0.44$ (blue, upper, dotted), and $\uo=0.38$ (blue, lower, dashed). The black solid line represents the mass of RX J0648.0-4418 ($M=1.28\pm0.05 M_\odot$) and the grey region shows the $1\sigma$ error bars. These figures were adapted from \cite{Jain:2015edg} }\label{fig:WD}
\end{figure}

Reference \cite{Jain:2015edg} have studied white dwarfs in Vainshtein breaking theories using the above equation of state by solving both the GR (equation \eqref{eq:HSEGR}) and Vainshtein breaking (equation \eqref{eq:HSEVAINBREAK}) HSEEs. The mass-radius relation that they obtained is shown in the left panel of figure \ref{fig:WD}. A $\chi^2$ test was performed using the observed masses and radii of 12 white dwarfs taken from \cite{Holberg:2012pu} treating $\uo$ as a fitting parameter. The resultant bounds were $-0.18\le\Upsilon_1\le0.27$ at $1\sigma$ and $-0.48\le\uo\le0.54$ at $5\sigma$. For $\uo<0$ the effects of Vainshtein breaking are equivalent to strengthening gravity, which has the effect of lowering the Chandrasekhar mass as shown in the central panel of figure \ref{fig:WD}. The mass of the heaviest observed white dwarf therefore places a bound on negative values of $\uo$ since values too negative would predict that this object should have gone supernova. The heaviest observed white dwarf has a mass $M=1.37\pm0.01M_\odot$ \cite{Hachisu:2000nx}, which places the bound $\uo\ge-0.22$ \cite{Jain:2015edg}. 

A final bound can be found by considering rotating white dwarfs. If the white dwarf is rotating with angular frequency $\omega$ then the HSEE must be augmented by a centrifugal force
\begin{equation}
\frac{\dd P}{\dd r}=-\frac{GM(r) \rho(r)}{r^2}\left[1+\frac{\pi\uo r^3}{M(r)}\left(2\rho(r)+r\frac{\dd\rho(r)}{\dd r}\right)\right]+\rho(r)\omega^2 r.
\end{equation}
If at any point the pressure gradient is outward i.e. $\dd P/\dd r>0$ then the star is unstable and so we must require \cite{Jain:2015edg}
\begin{equation}
\uo>\left(\frac{\omega^2}{\pi G\rho}-\frac{M(r)}{\pi r^3\rho(r)}\right)\left(2+\frac{\dd\ln\rho}{\dd r}\right)^{-1}
\end{equation}
at every $r$. Note that the equality changes to an upper bound if $\dd\ln\rho/\dd r<-2$. For the simple case of constant density one recovers the bound of \cite{Saito:2015fza}, $\uo>-2/3$. The positive pressure contribution implies that there is a minimum stellar mass for given values of $\omega$ and $\uo$, and that the strongest bounds should come from the most rapidly rotating objects. The majority of white dwarfs are slowly rotating but some rapidly rotating objects have been observed, in particular, RX J0648.0-4418, which has a mass $M=1.28\pm0.05 M_\odot$ which rotates with a period of 13.2 s \cite{Mereghetti:2011bh}. Fixing $\omega$ to this value, \cite{Jain:2015edg} have scanned a range of $x_0$ for different values of $\uo$ to find the range of parameters where such a star can be stable. Their results are illustrated in the right hand panel of figure \ref{fig:WD}. Accounting for the error bars, only values of $\uo$ in the range $-0.59\le\uo\le0.50$ can successfully model this object. 

\subsection{Distance Indicator Tests}
\label{sec:DI}

Distance indicators have proved a highly constraining novel probe of theories that screen using the thin-shell mechanism. Distance indicators are a method of inferring the distance to a galaxy based on some proxy, for example, by measuring the apparent magnitude of a standard candle such as a type Ia supernova. Typically, the formula used to infer the distance is based upon empirical calibrations made locally or theoretical calculations. In the former case, the calibration has been performed in a screened environment and in the latter the calculations assume GR. Therefore, if one compares two distance estimates to the same galaxy, one sensitive to the theory of gravity and the other not, then the two will not agree if the galaxy is unscreened. The amount by which they agree therefore constrains the model parameters. In what follows, we will summarize how \cite{Jain:2012tn} used two different distance indicators, Cepheids and tip of the red giant branch (TRGB) stars, to constrain thin-shell models.

\begin{figure}
\centering
\includegraphics[width=0.8\textwidth]{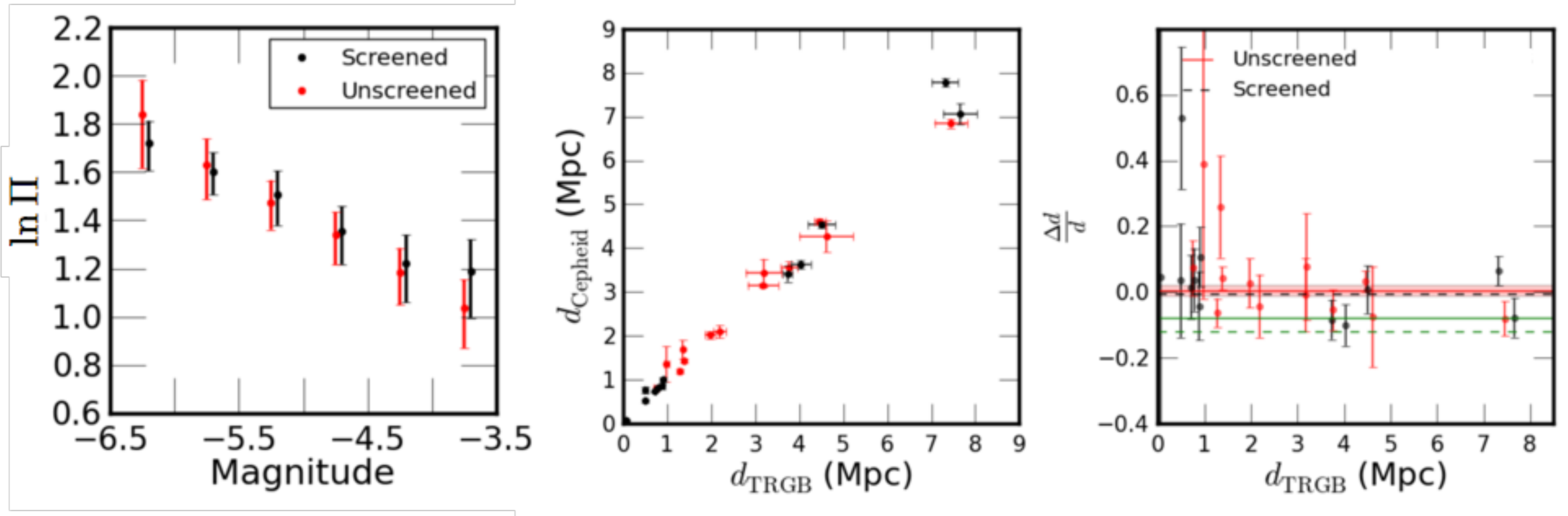}
\caption{Comparison of Cepheid and TRGB distances to a sample of screened (black dots) and unscreened galaxies (red dots). Figures adapted from \cite{Jain:2012tn}. \emph{Left Panel}: Comparison of the PL relation. \emph{Center Panel}: Comparison of Cepheid and TRGB distances. \emph{Right Panel}: Comparison of the difference between the two distance estimates as a function of the TRGB distance. The black dashed and red solid lines are the best-fit for the screened and unscreened samples respectively. The solid and dashed green lines show predictions for models with $2\beta^2(\phi_{\rm BG})=1$ and $1/3$ ($f(R)$ models) respectively.  }\label{fig:DI1}
\end{figure}

\subsubsection{Screened Distance Indicators: Tip of the Red Giant Branch}

When stars of $1$--$2M_\odot$ leave the main-sequence and ascend the red-giant branch the stellar luminosity is due to a thin shell of hydrogen burning outside the helium core. As the star ascends, the core temperature increases until it is hot enough for the triple-$\alpha$ process to proceed efficiently. At this point, known as the \emph{helium flash}, the star moves rapidly onto the asymptotic giant branch (AGB), leaving a visible discontinuity in the I-band at $I=4.0\pm0.1$ with the spread being due to a slight dependence on metallicity. This discontinuity can be used as a distance indicator since the luminosity is known. The details of the helium flash depend on nuclear physics and not the theory of gravity so the TRGB is a screened distance indicator\footnote{In fact, if $\chi_{\rm BG}\gsim10^{-6}$ MESA simulations reveal that the tip luminosity can decrease by 20\%. This is because the core is unscreened in these cases and the temperature is increased. For this reason, the temperature needed for the helium flash is reached faster and therefore the discontinuity occurs lower on the red giant branch. In what follows we will only consider $\chi_{\rm BG}\lsim 10^{-6}$.}.  

\subsubsection{Unscreened Distance Indicators: Cepheid Stars}

Stars with masses $3.5$--$10M_\odot$ execute semi-convection-driven \emph{blue loops} in the color-magnitude diagram where the temperature increases at roughly fixed luminosity. During this phase, the stars can cross the instability strip where they are unstable to pulsations driven by the $\kappa$-mechanism (see \cite{Cox1980} for details of this process). In this phase, a layer of doubly ionized helium acts as a dam for energy so that small compressions of the star go towards increasing the temperature in the ionization zone and not into increasing the outward pressure. This energy dam drives pulsations which result in a periodic variation of the luminosity and gives rise to a period-luminosity (PL) relation \cite{Freedman:2010xv}. 
These stars are known as \emph{Cepheid variable stars} and are used as distance indicators. In thin-shell screened theories, the inferred distance depends on the level of screening because the period of pulsation $\Pi$ is faster. This can be calculated either by solving the MLAWE \eqref{eq:MLAWE} or by using the fact that $\Pi\propto G^{-1/2}$ to find \cite{Jain:2012tn}
\begin{equation}\label{eq:DDD}
\frac{\Delta d}{d}=-0.3\frac{\Delta G}{G}\approx -0.6\beta^2(\phi_{\rm BG})\left(1-\frac{M(\rs)}{M}\right).
\end{equation}
Thus, in thin-shell screened theories Cepheid distance indicators are unscreened and under-estimate the true distance.

\subsubsection{Comparisons and Constraints}

\begin{figure}
\centering
\includegraphics[width=0.5\textwidth]{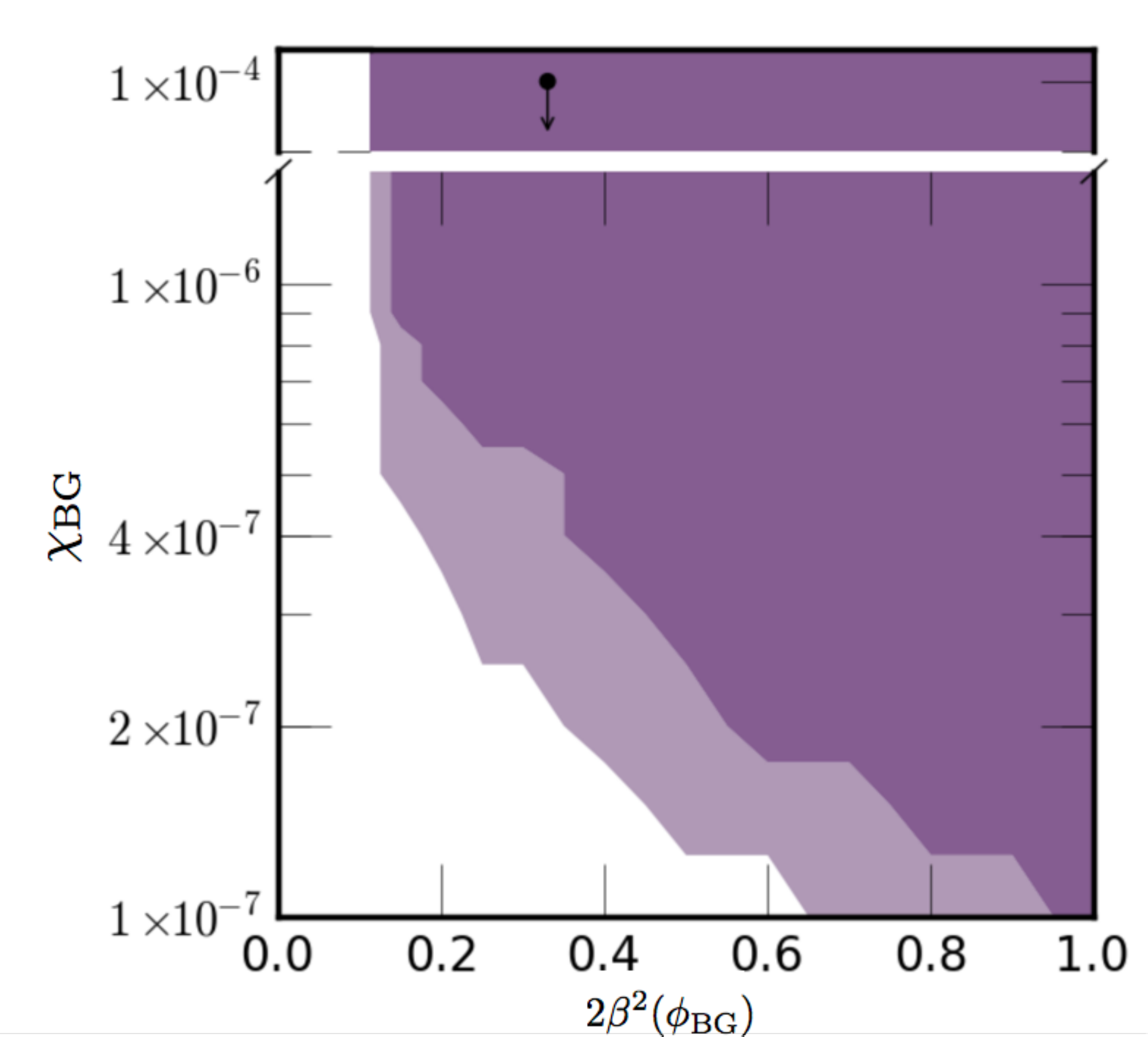}
\caption{The 68\% (dark purple) and 95\% excluded regions in the $\chi_{\rm BG}$--$\beta(\phi_{\rm BG})$ plane by comparing Cepheid and TRGB distance indicators. The black arrow shows an older constraint coming from galaxy cluster statistics. Figure adapted from \cite{Jain:2012tn}.  }\label{fig:DIcons}
\end{figure}

Using the screening map, \cite{Jain:2011ji} compared the TRGB and Cepheid for a sample of screened galaxies as well as a control sample of unscreened galaxies. The TRGB distance was taken as the true (screened) distance and the theoretical value of $\Delta d/d$ was computed by using MESA Cepheid profiles at the blue edge of the instability strip\footnote{The location of the instability strip may change in MG models but, to date, this has never been investigated.} to calculate $\Delta G/G$ in equation \eqref{eq:DDD}. An example is shown in figure \ref{fig:DIcons}. One can see that the two samples are consistent and a statistical analysis yielded the constraints shown in figure \ref{fig:DIcons}. In this case $\chi_{\rm BG}$ probes the cosmological value of $\chi$ (or, equivalently $f_{R0}$ for $f(R)$ models) since the galaxies are unscreened.  The bounds are the strongest astrophysical ones to date and $f_{R0}>4\times10^{-7}$ is ruled out for $f(R)$ models.

\subsection{Astroseisemology}
\label{sec:AS}
\begin{figure}
\centering
\includegraphics[width=0.7\textwidth]{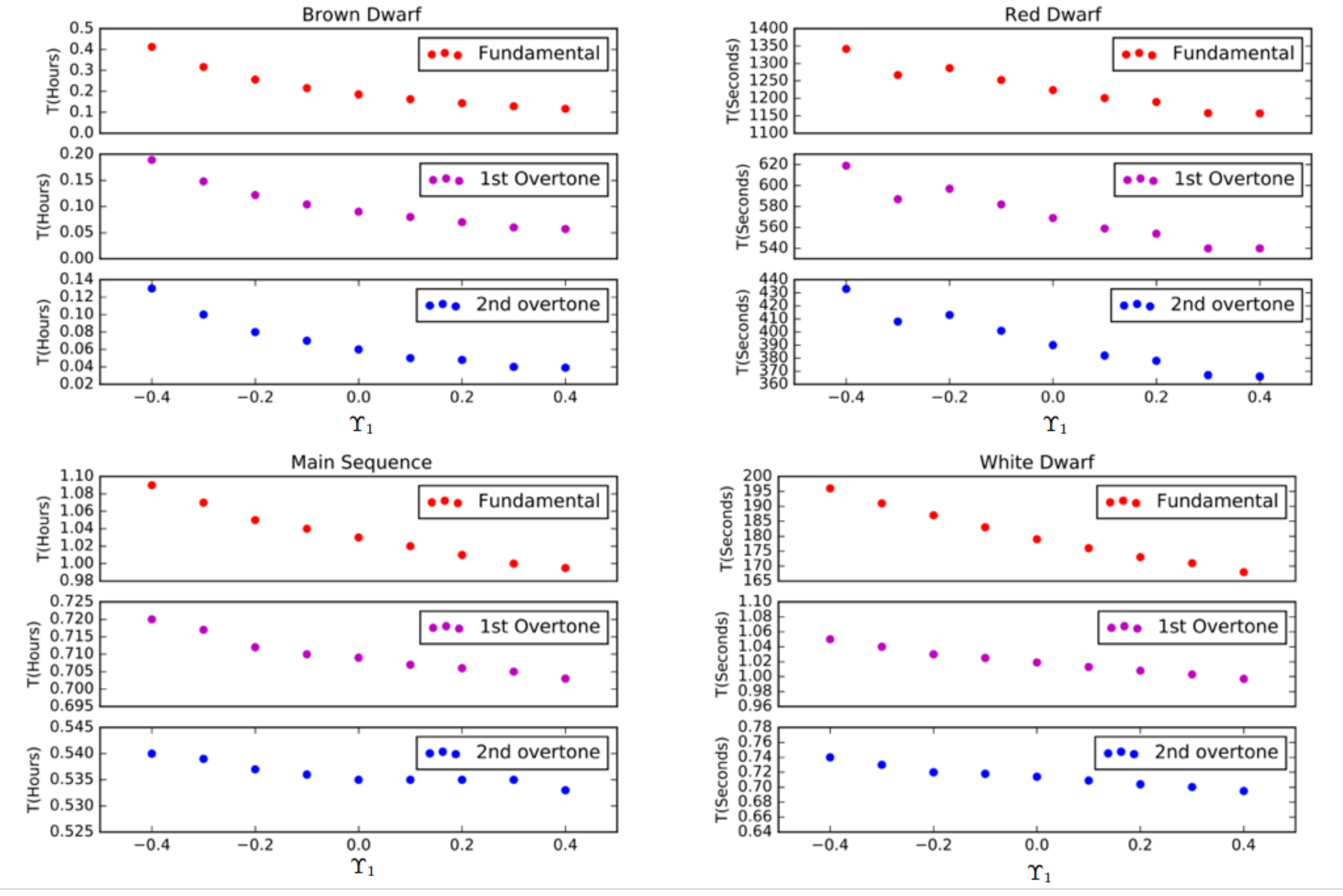}
\caption{The fundamental oscillation mode and first two overtones for representative polytropic stellar models in Vainshtein breaking theories. Figure adapted from \cite{Sakstein:2016lyj}.}\label{fig:VB_pulse}
\end{figure}

The use of radial stellar oscillations in Vainshtein breaking theories has been studied by \cite{Sakstein:2016lyj} who solved the MLAWE (equation \eqref{eq:MLAWEVB}) for some simple polytropic stellar models with the results shown in figure \ref{fig:VB_pulse}. The effects are small with the exception of brown dwarfs where $\Delta \Pi/\Pi\sim\mathcal{O}(1)$. The authors also investigated MESA models and found large changes in the period of Cepheid pulsations, although this was primarily driven by the altered equilibrium structure, which changed the intersection of the Hertzprung-Russell track with the instability strip.

\section{Galactic Tests}

\label{sec:galtests}

The morphology and dynamics of galaxies, in particular dwarf or low surface brightness (LSB) galaxies, have proved to be a strong tool for testing screened MG theories, especially those that screen using the thin-shell mechanism. This is partly because they have multiple components---dark matter, stars, gas---that can be screened to different levels and partly because they themselves have Newtonian potentials of $\mathcal{O}(10^{-8})$ making them some of the most unscreened objects in the universe\footnote{Of course, one must use dwarf galaxies that are sufficiently isolated so as to avoid environmental screening by their neighbors. In practice, this means using dwarf galaxies in voids. See the discussion in section \ref{sec:wheretolook} for more information on this matter.}. In this section, we discuss several novel tests that can, and in some cases have, been used to constrain thin-shell screening theories. We will use two common models for the dark matter density profile to aid in our computations: the cored isothermal sphere (CSIS)
\begin{equation}\label{eq:CSIS}
\rho(r)=\frac{\rho_0}{1+\left(\frac{r}{r_0}\right)^2}
\end{equation}
and the Navarro-Frenk-White (NFW) profile \cite{Navarro:1995iw}
\begin{equation}\label{eq:NFW}
\rho(r)=\frac{\rho_s}{\frac{r}{r_s}\left(1+\frac{r}{r_s}\right)^2}
\end{equation}
where $\rho_0$ and $r_0$ are the core density and radius (CSIS) and $\rho_s$ and $r_s$ are the scale density and radius (NFW). The former profile is typically a good fit to dwarf galaxies with core radii of order $1$--$4$ kpc \cite{Swaters2011} whilst the latter are well-motivated both theoretically and observationally.

\subsection{Rotation Curves}

\label{sec:DGcurves}

Theories that violate the equivalence principle i.e. those that screen using the thin-shell effect allow for a novel test of gravity using the rotation curves of different galactic components \cite{Jain:2011ji}. In particular, a galaxy is composed of stars (with Newtonian potentials $\pn\sim \mathcal{O}(10^{-7}$--$10^{-6})$) and diffuse gas (with Newtonian potential $\mathcal{O}(10^{-11}$--$10^{-12})$ \cite{Hui:2009kc}) that rotate around the center with a radially-dependent circular velocity given by
\begin{equation}\label{eq:vcirc}
\frac{v_{\rm circ}^2}{r}=\frac{G M_{\rm gal}(r)}{r^2}+\frac{Q}{M_{\rm obj}\mpl}\frac{\dd\phi_{\rm gal}}{\dd r},
\end{equation}
where a subscript `gal' refers to fields sourced by the galaxy and $M_{\rm obj}$ and $Q_{\rm obj}$ are the mass and scalar charge of the object respectively (see section \ref{sec:EP}). Let us make two simplifying assumptions: that the galaxy is unscreened so that $\dd\phi_{\rm gal}/\dd r=2\beta(\phi_{\rm BG})M(r)/\mpl$ and that stars are fully screened ($Q_{\rm obj}=0$) whilst the gas is fully unscreened $Q=\beta(\phi_{\rm BG}) M_{\rm obj}$. In this case, the circular velocity for the stars, $v_\star$ and gas $v_{\rm gas}$ satisfy \cite{Jain:2011ji}
\begin{equation}\label{eq:vcircratio}
\frac{v_{\rm gas}}{v_{\star}}=\sqrt{1+2\beta^2(\phi_{\rm BG})}.
\end{equation}
Thus, a comparison of the rotation curves of stars and gas can provide a novel probe of thin-shell screening theories. 

In practice, performing this test is not so simple because traditional probes of the galactic rotation curves use either H$\alpha$ or 21cm lines, both of which probe the gaseous, unscreened component. Another useful line is the OIII line that results from a forbidden transition in doubly ionized oxygen. This is particularly useful for thin-shell screening theories since the line is only present at very low densities. The stellar component can be probed independently using absorption lines for metals found in stellar atmospheres, for example, the MgI\emph{b} triplet or the CaII lines, found in the atmosphere of K- and G-dwarfs (main-sequence stars). These stars have Newtonian potentials of order $10^{-6}$ and hence values of $\chi_{\rm BG}$ smaller than this (where they are screened) can be probed provided that their host galaxies are unscreened for the same parameters. 

\begin{figure}
\centering
\includegraphics[width=0.5\textwidth,angle=-90]{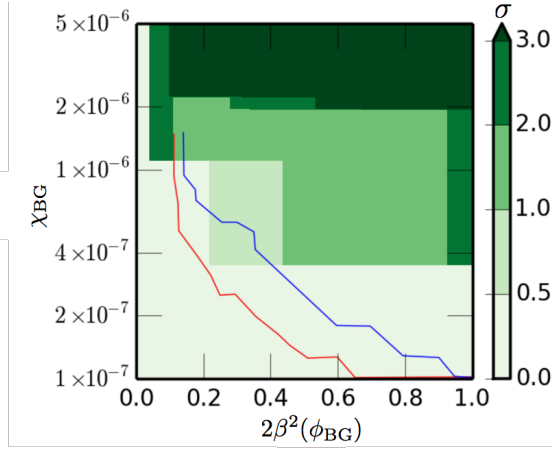}
\caption{Constraints on the value of $\chi_{\rm BG}$ and $\beta(\phi_{\rm BG})$ resulting from comparing the stellar and gaseous rotation curves of six unscreened galaxies in the screening map. The shade of green indicates the significance ($\sigma$) with which the models can be rejected; the exact scale is indicated in the figure. The $1$ and $2\sigma$ bounds from distance indicator tests described in section \ref{sec:DI} are shown in blue and red respectively. Figure adapted from \cite{Vikram:2014uza}.  }\label{fig:rot_curve}
\end{figure}

The screening map contains six galaxies that have both OIII and MgI\emph{b} information available that reference \cite{Vikram:2014uza} have used to perform this test. Their method is as follows: first, the gaseous rotation curve is used to fit a density profile for the galaxy accounting for systematic errors and astrophysical scatter. (Note that the gaseous curve is measured at more finely-spaced radial intervals so this provides amore accurate fit.) Next, this model is used to predict the stellar rotation curve and deviations from the measured curve are quantified to determine the statistical significance with which any deviation can be rejected. The results for each individual galaxy are then combined to obtain the constraints in figure \ref{fig:rot_curve}. (Note that these constraints probe the self-screening parameter ($\chi_{\rm BG}=\chi_0=3f_{R0}/2$) at cosmic densities since the galaxies are unscreened.) Also shown are the distance indicator constraints for comparison. One can see that distance indicators are more constraining for large couplings but rotation curves can push into the regime $2\beta^2(\phi_{\rm BG})<0.1$. (Effects on distance indicator tests are subdominant to GR in this range.) The jaggedness of the contours is a result of the small sample size. A larger sample size with better kinematical data from both gaseous and stellar emission lines would greatly improve the constraints. It is possible that data from SDSS IV-MaNGA could provide such a sample although, to date, no analysis has been performed.  

\subsection{Morphological and Kinematical Distortions}

\label{sec:DGmorph}

Another consequence of the WEP violations discussed in section \ref{sec:EP} is that when $\chi_{\rm BG}\lsim10^{-6}$ (the Newtonian potential of main-sequence stars) it is possible for the stellar component of a dwarf galaxy to be self-screening whilst the surrounding dark matter halo and gaseous component is unscreened. This leads to several novel morphological and kinematical tests of thin-shell screened theories \cite{Jain:2011ji}. If a galaxy of mass $M_1$ is falling edge-on towards another larger (but unscreened) galaxy of mass $M_2$ a distance $d$ away then the gas and dark matter will feel a larger external force than the stars and will hence fall at a faster rate. The stellar disk will then lag behind the gas and dark matter and become offset from the center. In the case of face-on infall the stars are displaced from the equatorial plane by a height \cite{Jain:2011ji}
\begin{equation}
z = \frac{2\beta^2(\phi_{\rm BG})M_1 R_0^3}{G M_2(R_0) d^2},
\end{equation}
where $R_0$ is the equilibrium distance from the galaxy's center. ($z$ and $R_0$ can be taken to define cylindrical coordinates centered on the falling galaxy.) Since $M_2\propto R_0^2$ with $n<3$ for any sensible density profile this is an increasing function of distance from the center and one hence expects the stellar disk to be warped into a U-shape that curves away from the direction of in-fall. 

\begin{figure}
\centering
\includegraphics[width=0.5\textwidth]{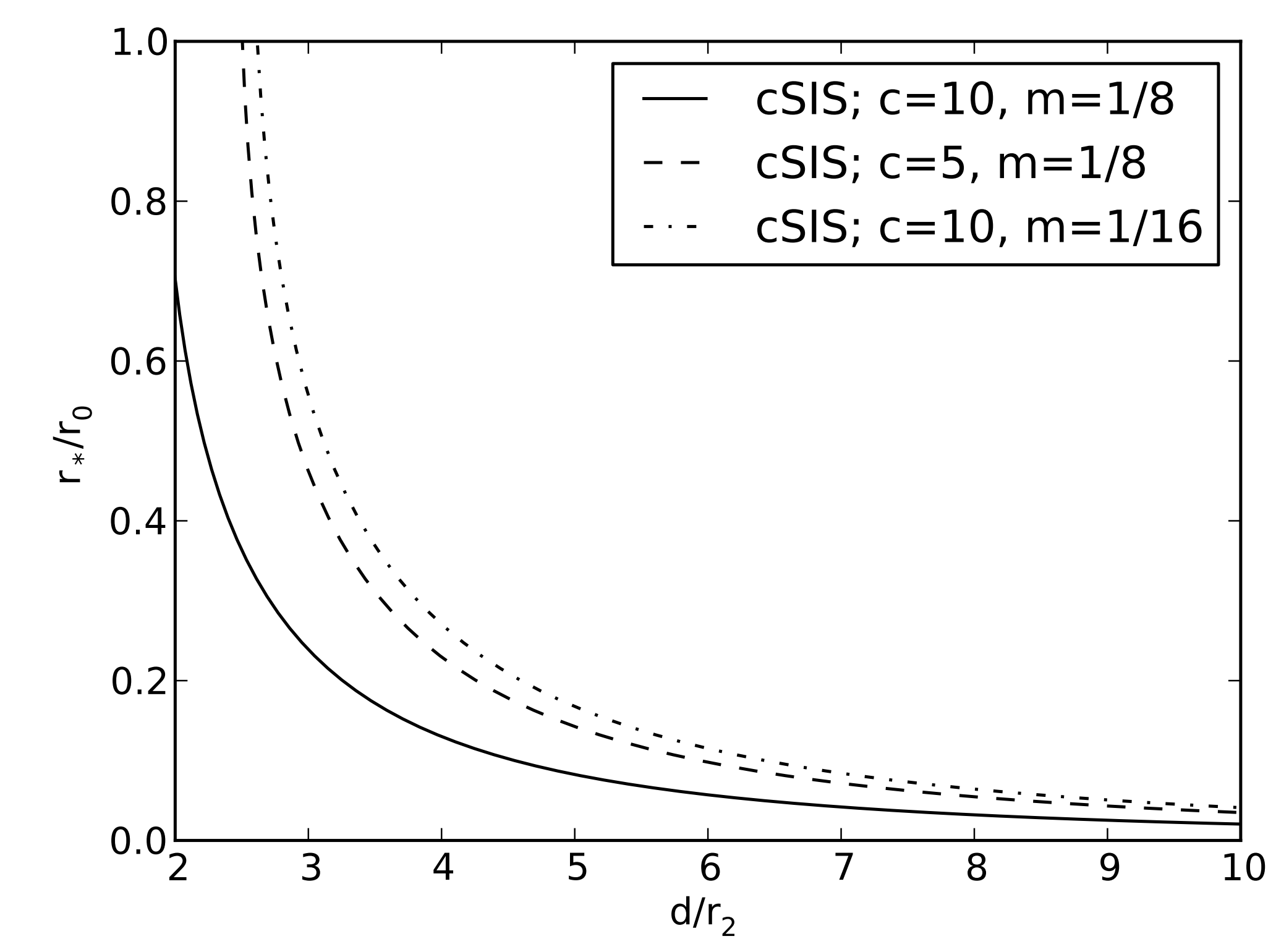}
\caption{The offset of the stellar disk as a function of $d/r_2$ ($r_2$ is the Virial radius of the falling galaxy) for CSIS profiles with parameters indicated in the figure. The figure uses concentrations $c=r_0/r_{\rm vir}$ with $r_{\rm vir}$ being the Virial radius and $m=M_2/M_1$. Figure taken from \cite{Jain:2011ji}.}\label{fig:DG1}
\end{figure}

Reference \cite{Jain:2011ji} have simulated these scenarios by solving for the orbits of galaxies composed of 4000 stars for dark matter halos described using both NFW and CSIS profiles. The halo and gas are taken to be fully unscreened with $\beta(\phi_{\rm BG})=1/\sqrt{2}$ (corresponding to a fifth-force that is equal in strength to the Newtonian force). The halo falls from a distance of 240 kpc to a final distance of 100 kpc in 3 Gyr. The orbits are initially circular with a gaussian scatter of 1 km/s. They considered two simple scenarios: edge-on infall and face-on infall. They identify the following three observational consequences of the WEP violation:

\begin{figure}
\centering
\includegraphics[width=0.6\textwidth]{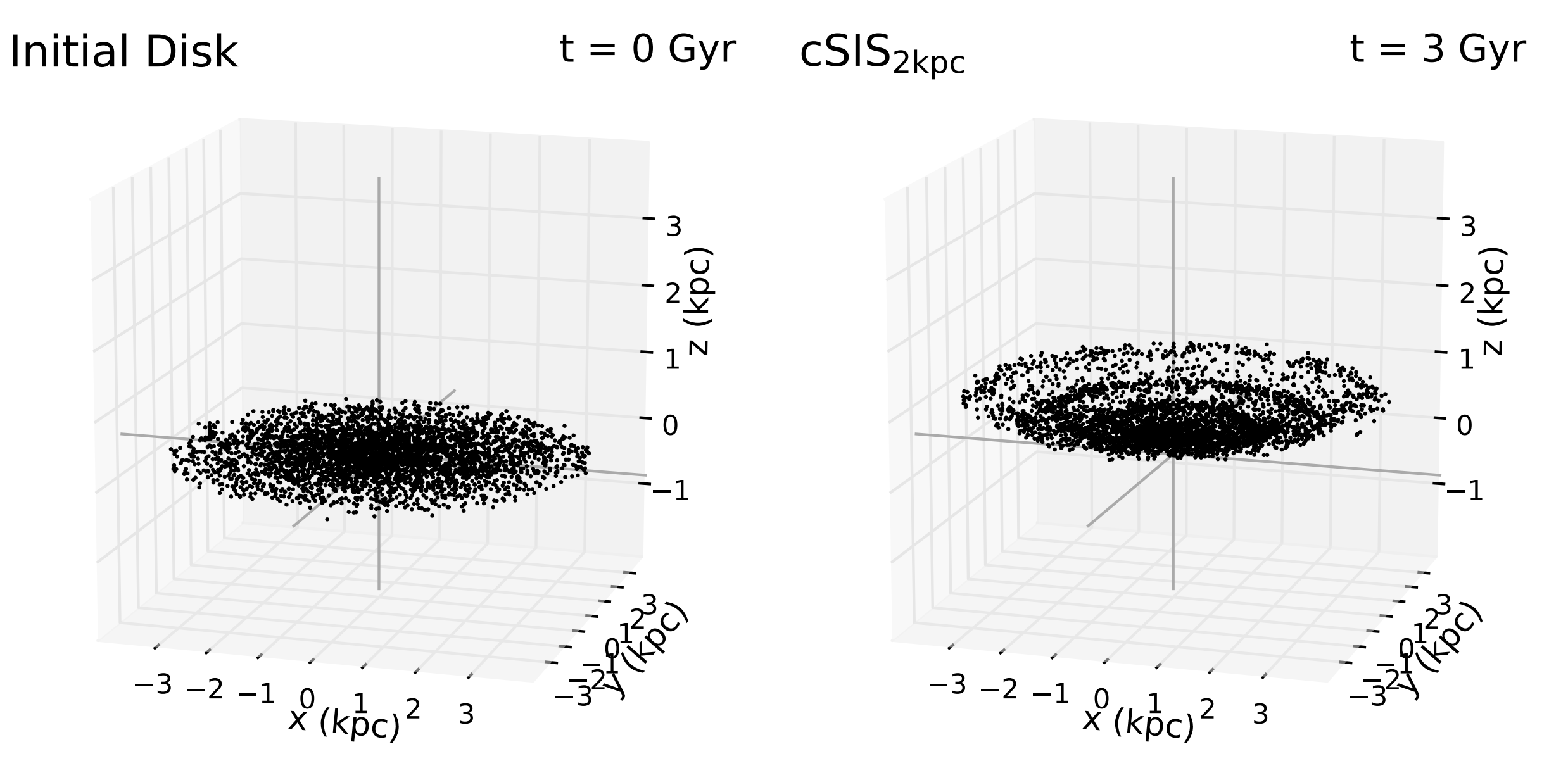}
\caption{Face-on infall, the neighboring galaxy lies in the negative $z$ direction. \emph{Left Panel}: The initial stellar distribution in the galaxy. \emph{Right Panel}: Final distribution for a CSIS profile with $r_0=2$ kpc and $\rho_0=1.2\times10^7M_\odot$/kpc$^3$. Figure adapted from \cite{Jain:2011ji} where more examples can be found.  }\label{fig:warping}
\end{figure}

\begin{figure}
\centering
\includegraphics[width=0.6\textwidth]{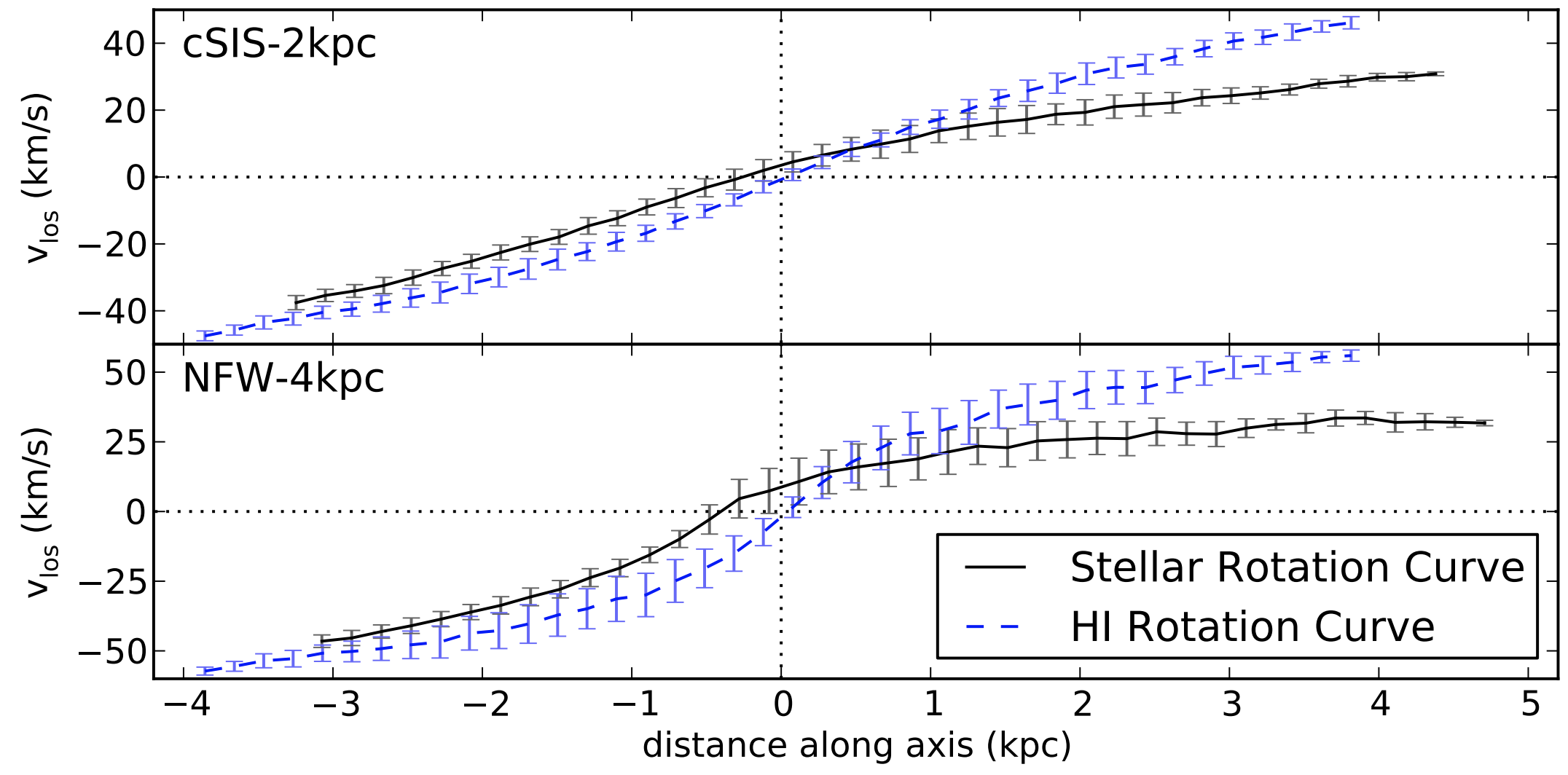}
\caption{The rotation curves ($v_{\rm los}$ is the line of sight velocity) for the stellar component (black, solid) and gaseous HI component (blue, dashed) after edge-on infall. The upper panel shows the same galaxy as figure \ref{fig:warping} and the lower panel shows an NFW profile with $r_s=4$ kpc and $\rho_s=10^7M_\odot$/kpc$^3$. Figure taken from \cite{Jain:2011ji}.  }\label{fig:asymmetry}
\end{figure}

\begin{itemize}
\item {\bf Offset stellar disk}: The reduced force on the stellar disk causes it to lag behind the HI gaseous component of the galaxy. An example of this is shown in figure \ref{fig:DG1} where $\mathcal{O}(\textrm{kpc})$ offsets are evident for CSIS galaxies. The offset is smaller for NFW profiles owing to the larger slope near the center and therefore larger restoring force. 
\item {\bf Morphological warping}: The face-on infall cases exhibit a warping of the galactic disks whereby the stars were displaced from the principal axis by an amount that increases with distance from the galactic center. An example of this is shown in figure \ref{fig:warping}.
\item {\bf Asymmetries in the rotation curves}: For edge-on in-falling galaxies, the stellar rotation curve becomes asymmetric compared with the HI curve. An example of this is shown in figure \ref{fig:asymmetry}. One can see that the zero-velocity point of the stellar rotation curve is off-axis (in the opposite direction to the galaxy's motion) whilst the HI curve is symmetric and sits on-axis. Note that the effect discussed in the previous section---faster HI circular velocities than stellar circular velocities due to self-screening of the stars and unscreening of the dwarf galaxy---is also evident in the plot. 
\end{itemize}

All of the effects found above are observable and the first attempt to use them to place constraints was made by reference \cite{Vikram:2013uba} who analyzed data circa 2013. They searched for potential offsets between the HI and optical centroids using SDSS r-band optical measurements to trace the stellar centroid and ALFALFA radio observations of the 21cm line to trace the HI centroid. In both cases they used a sample of unscreened galaxies taken from the screening map as well as a control sample of screened galaxies. A similar test was performed by looking for offsets between the optical centroid and kinematic HI centroid measured using the rotation curve. 
Both samples were consistent and a statistical analysis accounting for both astrophysical and MG scatter did not allow the authors to place any meaningful constraints. The same authors searched for U-shaped warpings of nearly edge-on galaxies by aligning each galaxy image so that the principal axis lies along the horizontal direction then finding the centroids in each vertical column; no constraints could be placed due to the large error bars. The authors estimate that 8000 dwarf galaxies would be needed to test down to $\chi_{\rm BG}\sim 10^{-6}$ and 20000 to reach $10^{-7}$. Finally, the authors tested the prediction of asymmetric rotation curves by using a weighted average of the difference in the velocity $\Delta v$ of the approaching and receding sides of the rotation curves for H$\alpha$ about the optical (stellar centroid) normalized to the maximum rotation velocity $v_{\rm max}$. The GHASP H$\alpha$ survey was used for this purpose. No constraints could be placed due to large uncertainties in the modeling of the inner halo as well as systematic uncertainties due to asymmetric drift and non-circular motion. 

Very recently, reference \cite{Desmond:2018euk} have used ALFALFA observations of a sample of 10,822 galaxies taken from an updated screening map \cite{Desmond:2017ctk} to constrain thin-shell theories by searching for offsets between the optical and HI centroids. Using a forward-modeling Bayesian likelihood method, they were able to obtain a new bound $\chi_0=3f_{R0}/2<1.5\times10^{-6}$. Improved measurements and larger samples from future surveys such as VLA or SKA could markedly improve these constraints. In particular, one could constrain $\beta(\phi_{\rm BG})\lsim 10^{-3}$ \cite{Desmond:2018euk}.

    

\section{Galaxy Cluster Tests}

\label{sec:clustertests}

Galaxy clusters are another useful probe of MG models. One reason for this is that they can be probed using both non-relativistic (dynamical and kinematic) and relativistic (weak lensing) tracers and many MG theories predict that the dynamical and lensing masses differ. Another is that they are some of the most massive objects in the universe and may enhance small fifth-forces (although they are also likely to be highly self-screening).

\subsection{Dynamical vs. Lensing Mass: X-ray and Lensing Comparisons}

\label{sec:xray}

There are many ways to probe the mass of galaxy clusters. Dynamical measurements such as rotational velocities or the surface brightness temperature use non-relativistic objects such as the galaxies themselves or intra-cluster medium gas whilst weak lensing provides a relativistic probe. In GR, the mass measured using both types of tracer will agree but in generic MG theories the dynamical mass (measured using non-relativistic objects) and lensing mass will differ. Typically, one can quantify this difference using the PPN parameter $\gamma$ but in screened theories this is close to unity in the solar system\footnote{In the case of Vainshtein breaking theories it is precisely unity \cite{Koyama:2015oma}.} and the deviation is a function of how screened the cluster is (see \cite{Burrage:2016bwy} and especially \cite{Burrage:2017qrf} section 3.2 for an extended discussion on this). In what follows, we will look at two tracers and two different definitions of mass.

The intra-cluster ionized plasma is in hydrostatic equilibrium and the pressure profile is therefore dependent on the dark matter mass (which is the dominant contribution to the gravitational force), which is well-modeled by an NFW profile \eqref{eq:NFW}. The gas emits in the X-ray and the surface brightness can be directly related to the pressure, allowing the mass profile to be determined. We therefore define the thermal mass via
\begin{equation}\label{eq:mtherm}
M_{\rm therm}=-\frac{r^2}{G\rho_{\rm gas}(r)}\frac{\dd P}{\dd r}.
\end{equation}
In theory, there is a component of non-thermal pressure that could act as a correction to this so that the thermal mass does not truly probe the non-relativistic source for gravity. This has hitherto been ignored and chameleon simulations have shown it to be negligible \cite{Wilcox:2016guw}. Using the ideal gas law, $P_{\rm gas}=k_{\rm B} n_{\rm gas} T_{\rm gas}$ where $n_{\rm gas}$ is the gas number density one has
\begin{equation}
M_{\rm therm}=-\frac{k_{\rm B}T_{\rm gas}(r) r}{\mu m_{\rm p}G}\left(\frac{\dd\ln\rho_{\rm gas}}{\dd\ln r}+\frac{\dd\ln T_{\rm gas}}{\dd\ln r}\right),
\end{equation}
where $\mu$ is the mean molecular weight and $m_{\rm p}$ is the proton mass so that $\rho_{\rm gas}=\mu m_{\rm p}n_{\rm gas}$. One can also relate the gas number density to the electron density via $n_{\rm e}=(2+\mu)n_{\rm gas}/5$. Using a combination of X-ray and SZ measurements, one can apply fitting functions to determine $T_{\rm gas}(r)$ (which is taken to be the electron temperature) and $n_{\rm e}(r)$ and therefore infer the thermal mass. See reference \cite{Terukina:2013eqa} Appendix A for further details.

The dynamics of light is controlled by the lensing potential $\Phi+\Psi$, which in GR satisfies $\nabla^2(\Phi+\Psi)=8\pi G\rho$. One integrate once to find $\Phi'+\Psi'=GM(r)/r^2$, which motivates the definition of a lensing mass
\begin{equation}\label{eq:Mlens}
M_{\rm lens}=\frac{r^2(\Phi'+\Psi')}{2G},
\end{equation}
which can be measured using the lensing shear. In GR, the thermal (dynamical) and lensing masses are identical and probe the dark matter component\footnote{In theory, all the components are probed but the dominant contribution is from the dark matter halo.} and so comparing the two is a novel probe of screened MG theories. 

\subsubsection{Thin-Shell Screened Theories}

\begin{figure}
\centering
\includegraphics[width=0.5\textwidth]{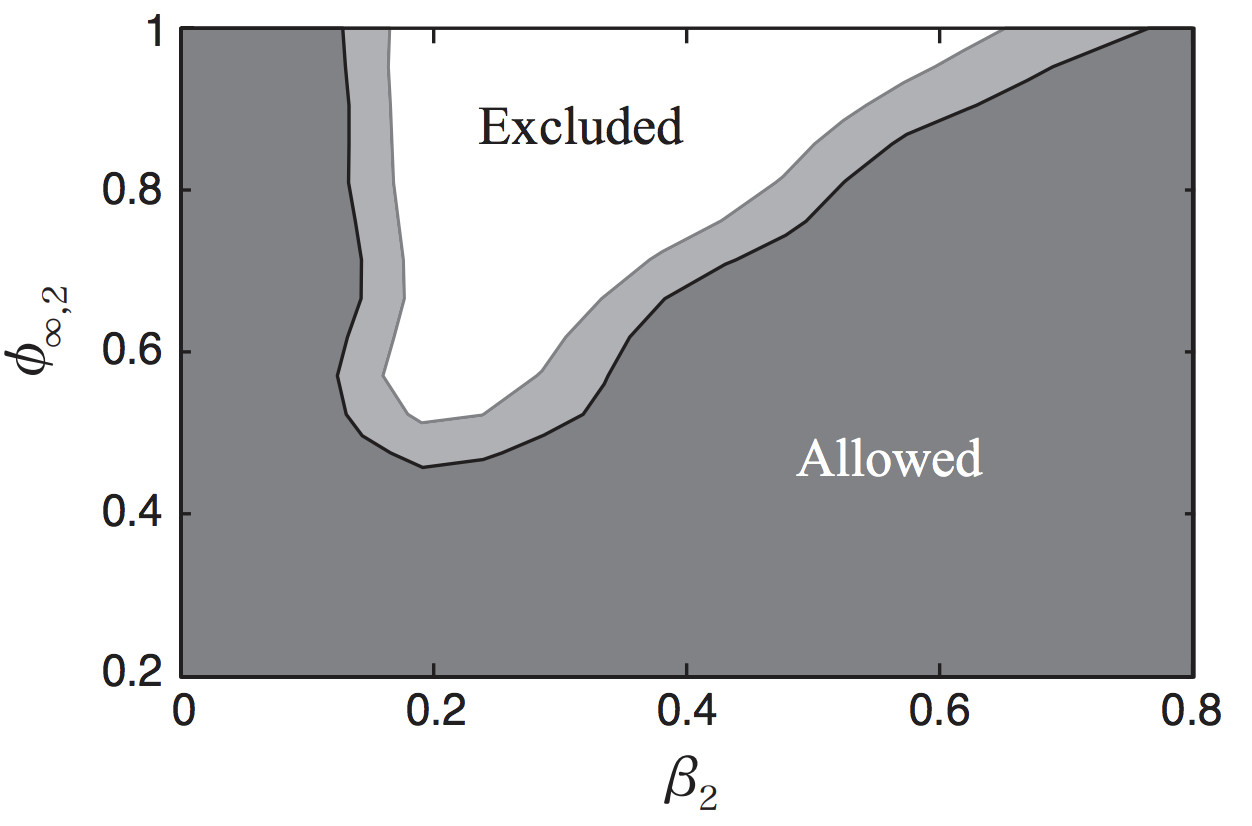}
\caption{Exclusion plot for chameleon theories coming from a comparison of the thermal and lensing mass of the Coma cluster. The 95\% (dark grey) and 99\%(light grey) confidence limits are shown. The quantity $\phi_{\infty,\,2}=1-\exp(-\phi_{\rm BG}/(10^{-4}\mpl))$ and $\beta_2=\beta(\phi_{\rm BG})/(1+\beta(\phi_{\rm BG}))$. Figure taken from \cite{Terukina:2013eqa}.}\label{fig:cham_lensing}
\end{figure}

Theories that screen using the thin-shell mechanism are conformal scalar-tensor theories and therefore the lensing of light is unaffected so that the lensing mass is the true mass\footnote{By this we mean the mass found by integrating over the dark matter density profile.} whereas the thermal (dynamical) mass is given by
\begin{align}\label{eq:mthermcham}
M_{\rm therm}(r)&= M(r)-\frac{r^2}{G}\frac{\beta(\phi_{\rm BG})}{\mpl}\frac{\dd\phi}{\dd r}
\end{align}
where $M(r)$ is the mass found by integrating the density profile. The second term is deviation from the lensing mass due to MG and \cite{Terukina:2013eqa} have placed constraints on chameleon models by using observations of the Coma cluster to constrain the deviation between this and the lensing mass. This was accomplished by assuming an NFW profile \eqref{eq:NFW} and using an analytical approximation for $\phi(r)$ (obtained using said profile).
Their constraints are shown in figure \ref{fig:cham_lensing} and the bound $\chi_0=3f_{R0}/2<9\times10^{-5}$ was obtained. A followup analysis was performed by \cite{Wilcox:2015kna} using a sample of 58 X-ray selected clusters for which both temperature data from XMM-Newton and lensing data from CFHTLens were available; similar constraints were obtained. Currently, these constraints are not competitive with those coming from distance indicators, although this may change in the future since the number of X-ray and lensing measurements are expected to increase. (Such measurements would be applicable to a diverse range of science goals.)

\subsubsection{Theories with Vainshtein Breaking }

In theories with Vainshtein breaking, both the thermal and lensing mass are altered since both $\Phi$ and $\Psi$ receive corrections. Assuming an NFW profile, the masses are given by\footnote{The thermal mass and lensing mass with $\Upsilon_3=0$ was first derived by \cite{Sakstein:2016ggl}. These expressions are fully general and are presented here for the first time.}
\begin{align}\label{eq:MthermVB}
M_{\rm therm}(r)&= M(r) + \pi \uo r_s^3\rho_s\left(1-\frac{r_s}{r}\right)\left(1+\frac{r_s}{r}\right)^{-3}\\
M_{\rm lens}(r) & = M(r) + \frac{\pi r_s^3}{2}\rho\left[\left(\uo+5\ut+4\Upsilon_3\right)-\left(\uo+5\ut+4\Upsilon_3\right)\frac{r_s}{r}\right]\left(1+\frac{r_s}{r}\right)^{-3}.
\end{align}
The case $\Upsilon_3=0$ was studied by \cite{Sakstein:2016ggl} who used the stacked profiles of the same X-ray selected sample used by \cite{Wilcox:2015kna} to place the new constraints
\begin{equation}
\label{eq:lensingconstraintsVB}
\uo=-0.11^{+ 0.93}_{-0.67}\textrm{ and } \ut=-0.22^{+1.22}_{-1.19}
\end{equation}
at the 95\% confidence level. Note that the mean redshift for the cluster sample is $z=0.33$ and since $\uo$ and $\ut$ can vary with time these constraints should be taken to apply at this redshift. Models where there is no strong time-dependence are constrained to this level at $z=0$.

\subsection{Strong Equivalence Principle Violations: Black Hole Offsets}
\label{sec:offsets}
\begin{figure}
\centering
{\includegraphics[width=0.45\textwidth]{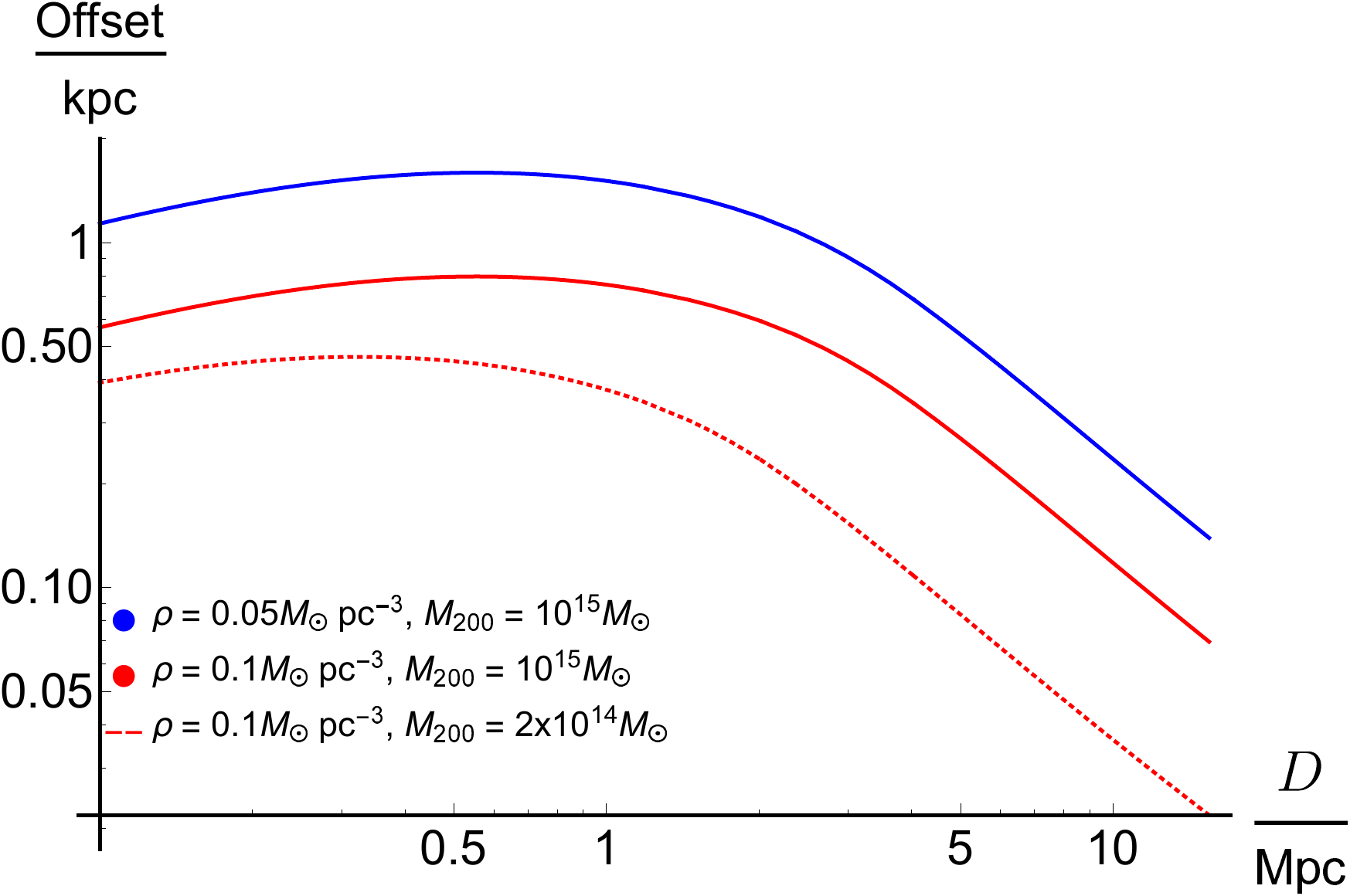}}
\caption{
The predicted offset for a cubic galileon model with $r_c=500$ Mpc as a function of distance from the Virgo cluster center for different satellite central densities and cluster masses ($M_{200}$ is the mass enclosed inside $R_{200}$) indicated in the figure. Figure taken from \cite{Sakstein:2017bws}.}\label{fig:Virgo}
\end{figure}

Galileon theories are difficult to test on small scales (unless they include Vainshtein breaking) due to the efficiency of the Vainshtein mechanism and, until recently, the strongest constraints came from the lack of deviations in the inverse-square law found using lunar laser ranging (LLR) \cite{Dvali:2002vf}. (Laser ranging to Mars could improve these by several orders of magnitude \cite{Sakstein:2017pqi}.) One way the Vainshtein mechanism has been successfully constrained is using the SEP violations discussed in section \ref{sec:SEP}. The principle of this test, as first pointed out by \cite{Hui:2012jb}, is the following: consider a galaxy falling in an external Newtonian and galileon field. The baryons (stars and gas) and dark matter all have scalar charge $Q=M$ but the central black hole (in fact, any black hole) has zero scalar charge. The stars, gas, and dark matter therefore fall faster than the central black hole, causing it to lag behind and become offset from the center. Eventually, the restoring force from the remaining baryons at the center will compensate for the lack of the galileon force, leading to a visible offset\footnote{In fact, it is possible for the black hole to escape the galaxy all together in some circumstances \cite{Sakstein:2017bws}.}. This offset would be correlated with the direction of the galaxy's acceleration, thereby providing a smoking-gun signal. 

\begin{figure}
\centering
{\includegraphics[width=0.45\textwidth]{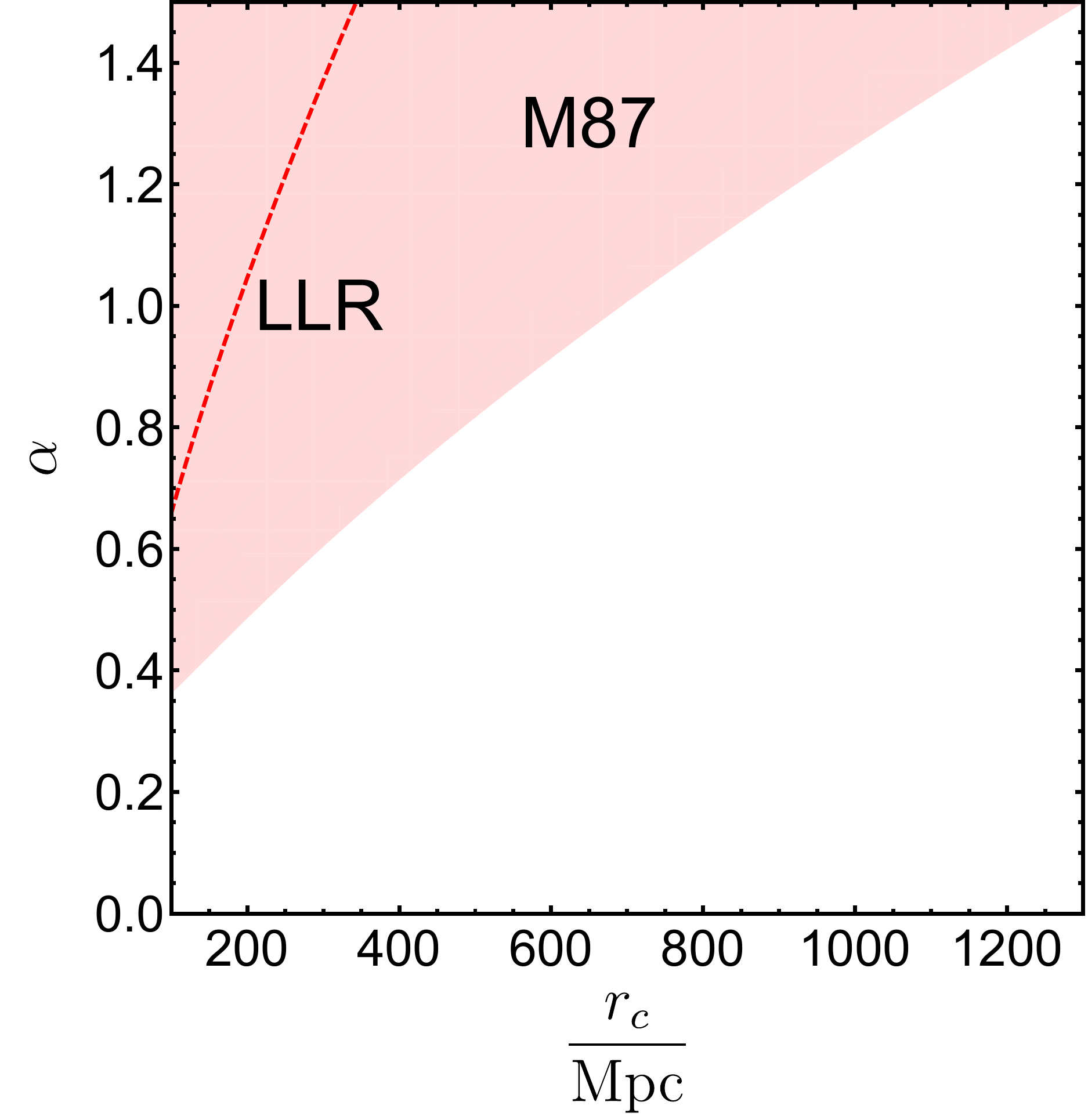}}
{\includegraphics[width=0.45\textwidth]{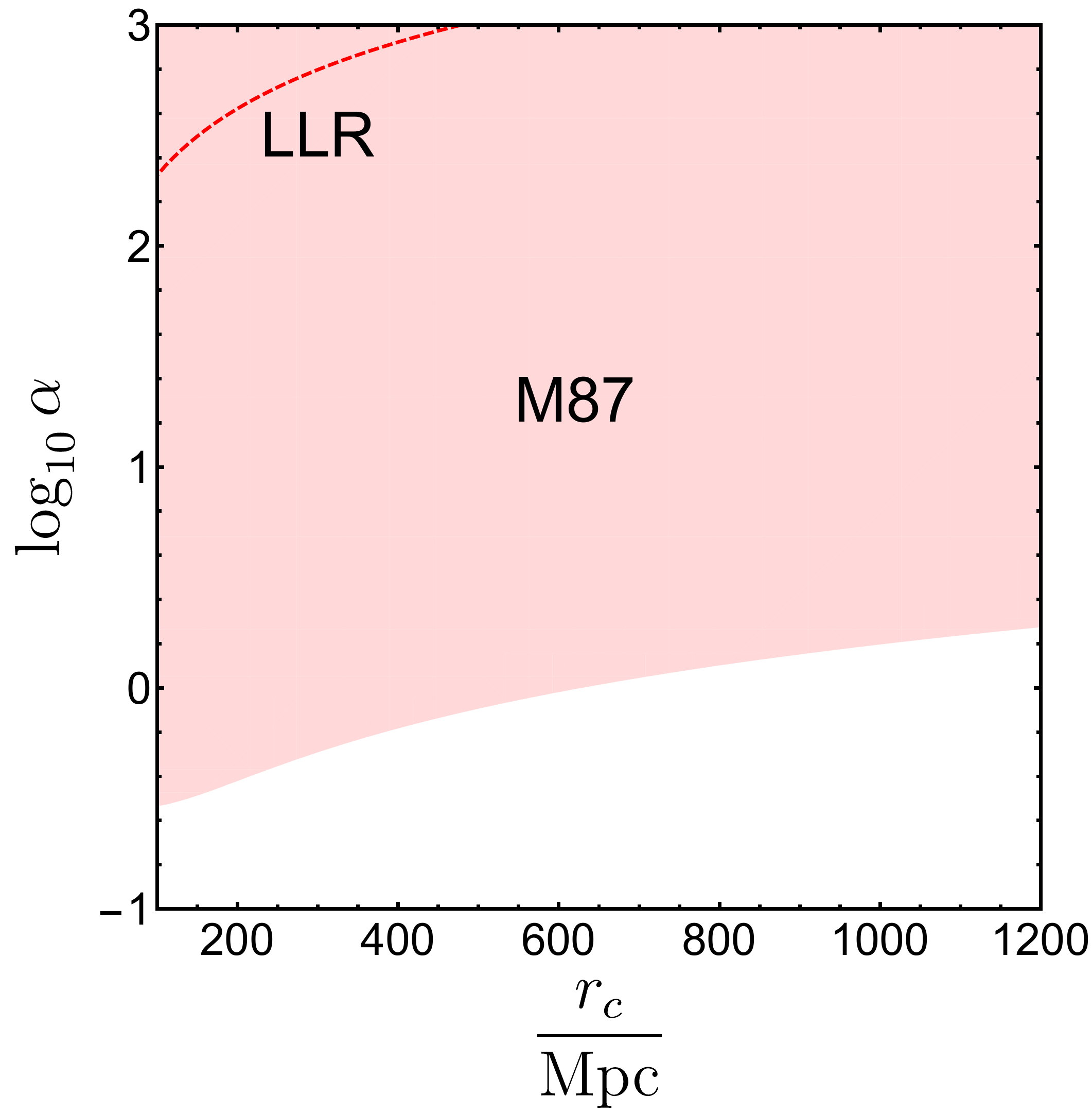}}
\caption{Constraints on the galileon model parameters from the lack of a black hole offset in M87 in the Virgo cluster (pink shaded region). The red dashed line shows the older constraints from LLR. \emph{Left Panel}: Cubic galileon \emph{Right Panel}: Quartic galileon. Note that the scale for $\alpha$ is logarithmic. Figure in left panel adapted from \cite{Sakstein:2017bws}.}\label{fig:gal_cons}
\end{figure}

One scenario for testing this (inspired by the proposal of  \cite{Hui:2012jb}) was proposed by \cite{Sakstein:2017bws}. Satellite galaxies orbiting inside massive clusters are accelerating towards the center. When they are far away they can be outside the Vainshtein radius and see an unscreened galileon field but even inside the Virial radius there can be a large galileon contribution to the acceleration. (This is partly because the Vainshtein mechanism is not as efficient for extended objects \cite{Schmidt:2010jr} and partly because 2-halo corrections boost the cluster mass at large radii \cite{Diemer:2014xya}.) 
Figure \ref{fig:Virgo} shows the predicted offsets for the Virgo cluster (modeled using a concentration $c=5$ NFW profile) for satellite galaxies with constant density profiles. One can see that offsets of $\mathcal{O}(\textrm{kpc})$ are predicted. Using a dynamical model of M87 \cite{Asvathaman:2015nna}, which is falling towards the center of the Virgo cluster, one finds that the central black hole is offset by no more than 0.03 arcseconds so that the galileon force $\lsim 1000 (\textrm{km/s})^2/\textrm{kpc}$. Combining this with the model for the Virgo cluster above, \cite{Sakstein:2017bws} obtained the constraints on cubic and quartic galileon models shown in figure \ref{fig:gal_cons}. One can see that self-accelerating models ($r_c\sim 6$--$10\times10^{3}$ Mpc) are currently unconstrained but smaller values of $r_c$ are excluded. Of course, this is just one system and \cite{Sakstein:2017bws} discuss how future X-ray and optical surveys could improve these bounds.

It is worth mentioning here that the black hole offset test is not unique to Vainshtein screened theories or, indeed, screened MG. Any scalar-tensor theory will predict similar SEP violations. What is novel is the screening mechanism. In the absence of Vainshtein screening, scalar-tensor theories are best tested in the laboratory or solar system\cite{Burrage:2016bwy,Burrage:2017qrf} (or with the other astrophysical probes discussed here in the case of thin-shell screening). Vainshtein screening is so efficient that this difficult test is the most competitive.  

\section{Relativistic Stars}

\label{sec:relSS}

Relativistic stars are a good probe of alternative gravity theories but many of the classic tests (absence of dipole radiation for example) are not competitive for screened MG. In the case of thin-shell screening, the screening is more efficient for objects with larger Newtonian potentials (i.e. relativistic objects) and any effects are highly degenerate with the EOS \cite{Brax:2017wcj}. In the case of Vainshtein screening, the Vainshtein radius is several orders of magnitude larger than the radius of neutron stars and any deviations from GR are highly suppressed \cite{Chagoya:2014fza}. The exception is theories with Vainshtein breaking since the deviations inside astrophysical bodies can be important for the structure of compact objects. Given the above considerations, the entirety of this section will focus on Vainshtein breaking theories. See \cite{Berti:2015itd,Babichev:2016rlq} for reviews of compact objects in MG theories. 

One generic feature of scalar tensor theories with coupling strength $\alpha$ ($=2\beta^2(\phi_{\rm BG})$ for chameleons) is that there is a tachyonic instability for the scalar when the quantity\footnote{This condition is equivalent to demanding that the trace of the energy-momentum tensor, which sources the scalar, becomes negative.}
\begin{equation}
1-3\frac{\tilde{P}}{\tilde{\rho}}\ge-\frac{1}{12\alpha^2},
\end{equation}
where tildes refer to Jordan frame quantities. This instability is never realized for screened modified gravity theories when sensible equations of state are used \cite{Babichev:2009td,Babichev:2009fi}.

\subsection{Vainshtein Breaking}

\label{sec:NS}

\subsubsection{Static Spherically Symmetric Stars}

Unlike non-relativistic objects, for which the HSEE depends universally on $\uo$ independently of the specific theory, the TOV equation for Vainshtein breaking theories depends on both the theory, and the asymptotics. Indeed, since $\uo$ is a function of the cosmological time-derivative of the scalar (as well as $H$ and second time-derivatives) one has $\uo=0$ in an asymptotically Minkowski spacetime whereas in an asymptotically FRW spacetime $\uo$ may be non-zero. The situation is complicated further by the fact that there are three branches of solution for the scalar field (the equation of motion reduces to a cubic after manipulation) and the correct branch (the one which gives the correct asymptotics) requires a fully relativistic calculation to determine. For this reason, current works have used specific models that admit exact de Sitter (dS) solutions so that one can determine the correct branch of solution (and therefore $\Upsilon_1$) in a controlled and systematic manner. References \cite{Babichev:2016jom,Sakstein:2016lyj,Sakstein:2016oel} have identified several models that have exact dS solutions and exhibit Vainshtein breaking. 
%
The derivation of the TOV equation for these models is long and complicated so we refer the reader to \cite{Babichev:2016jom,Sakstein:2016lyj,Sakstein:2016oel} for the full details. Here we will only sketch the derivation. One first solves the equations of motion to find an exact de Sitter solution. Next, the metric potentials and scalar are perturbed by introducing a perfect fluid source. One finds an exact Schwarzschild-de Sitter metric in the exterior of the star and the correct branch of solution for the scalar inside the star is the one that matches onto this solution. Taking the sub-Horizon limit, one can eliminate the scalar completely leaving a system of three equations that must be solved, two for the metric potentials and one for the pressure. These are the Vanshtein breaking counterparts to the Tolman-Oppenheimer-Volkoff (TOV) equation found in GR. Given an equation of state, these can be solved with appropriate boundary conditions (spherical symmetry at the center and vanishing pressure at the radius) to find the structure of the star. In the simplest models, one finds a universal parameter $\uo=\ut=\Upsilon$ ($\Upsilon_3=0$).

\begin{figure}
\centering
{\includegraphics[width=0.4\textwidth]{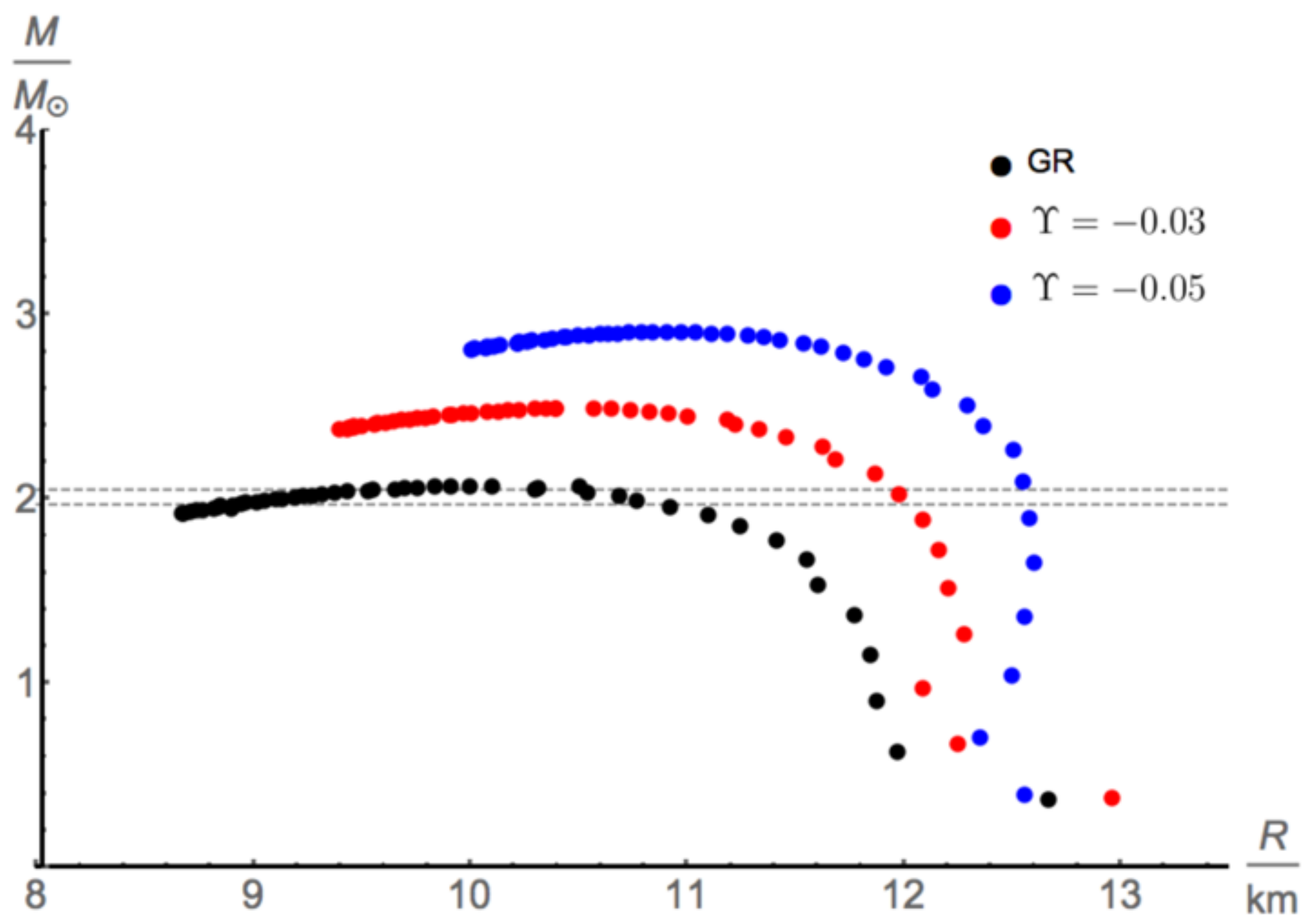}}
{\includegraphics[width=0.4\textwidth]{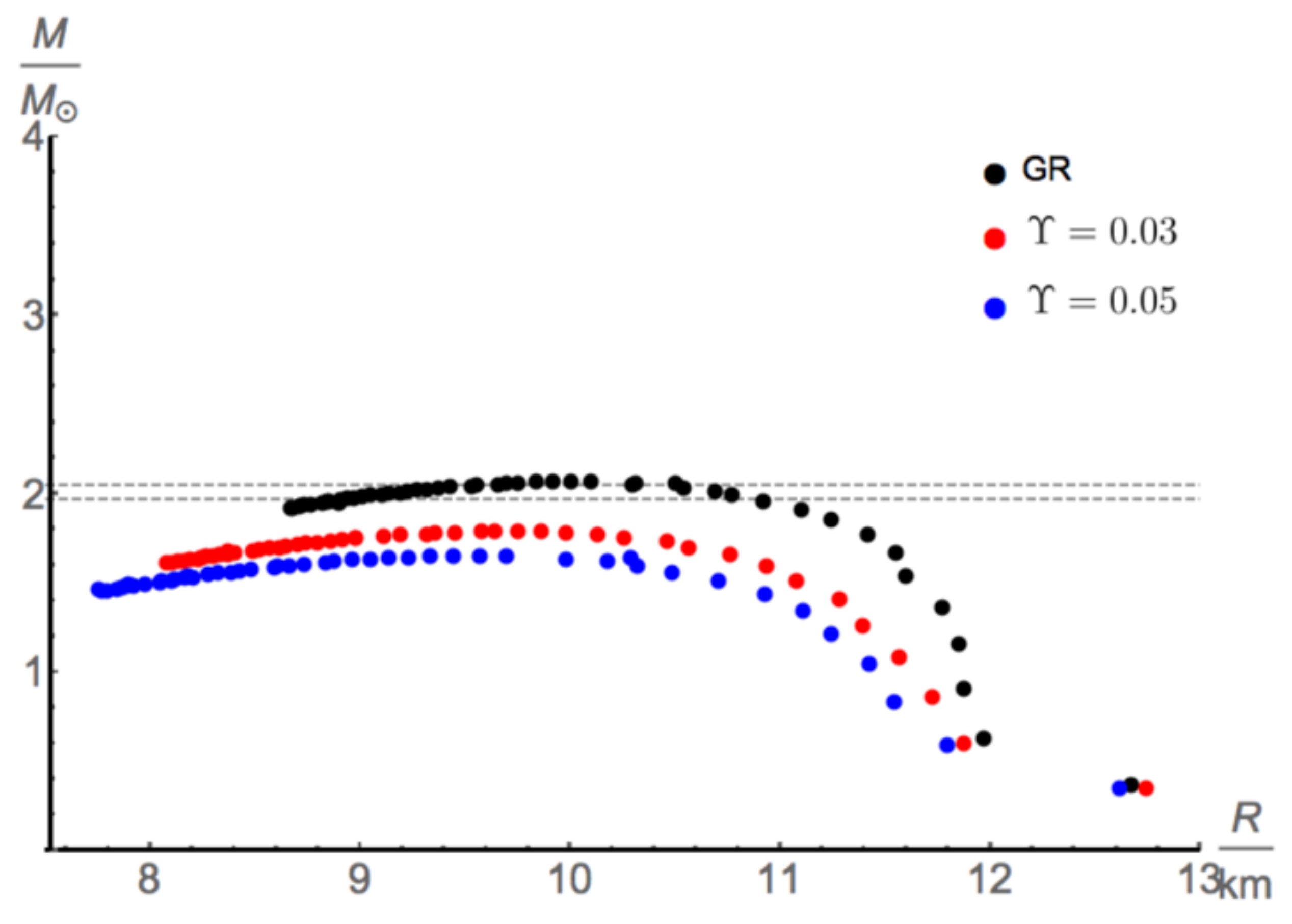}}
\caption{The mass-radius relation for neutron stars in Vainshtein breaking theories for the values of $\Upsilon$ given in the figures. The GR prediction corresponds to $\Upsilon=0$ and the grey dashed line shows the upper and lower bounds on the mass of the heaviest neutron star presently observed \cite{Demorest:2010bx}. Figures taken from \cite{Babichev:2016jom}. \emph{Left Panel}: $\Upsilon$<0. From top to bottom: $\Upsilon=-0.05$, $\Upsilon=-0.03$, GR. \emph{Right Panel}: $\Upsilon$>0. From top to bottom: $GR$, $\Upsilon=0.03$, $\Upsilon=0.05$. }\label{fig:VBrel1}
\end{figure}

Reference \cite{Babichev:2016rlq} has solved the TOV equations using both $n=2$ relativistic polytropic and two realistic (BSK20 and SLy4) neutron star equations of state. Furthermore, reference \cite{Sakstein:2016oel} have solved using 32 equations of state, including some that include hyperons, kaons, and strange quark matter. An example of the neutron star mass-radius relation found using the SLy4 EOS is shown in figure \ref{fig:VBrel1}. Also shown is the mass of the heaviest neutron star observe to date (PSR J0348+0432 with mass $M=2.01\pm0.04M_\odot$) \cite{Demorest:2010bx}. One can see that positive values of $\Upsilon$ make the stars less compact i.e. they have lower masses at fixed radii. In the case of SLy4, even reasonably small values of $\Upsilon>0$ result in maximum masses that are not compatible with the mass of PSR J0348+0432 but since the EOS of dense neutron matter is not known this just implies that the SLY4 EOS is not compatible with Vainshtein breaking theories with these values of $\Upsilon$ and, indeed, changing the EOS one can find masses in excess of $2M_\odot$. Negative values of $\uo$ produce stars that are more compact so that they have larger masses and fixed radii. Although figure \ref{fig:VBrel1} only shows one EOS, the features exhibited there are generic for all the equations of state studied by \cite{Sakstein:2016oel}. Another interesting prediction evident from the figure is that the maximum mass can be far in excess of the GR prediction. While an exact mass is dependent on the EOS, reference \cite{Sakstein:2016oel} has noted that, for some equations of state, the radius and mass of higher-mass stars can violate the GR causality bound, and so the observation of stars with these properties would be in tension with GR. The equivalent causality bound in Vainshtein breaking theories is unknown at present since calculating it is a more difficult task. In particular, the kinetic mixing of the scalar and metric would require one to find the sound speeds of both the scalar and pressure modes simultaneously.

\subsubsection{Slowly Rotating Stars}

\begin{figure}
\centering
{\includegraphics[width=0.45\textwidth]{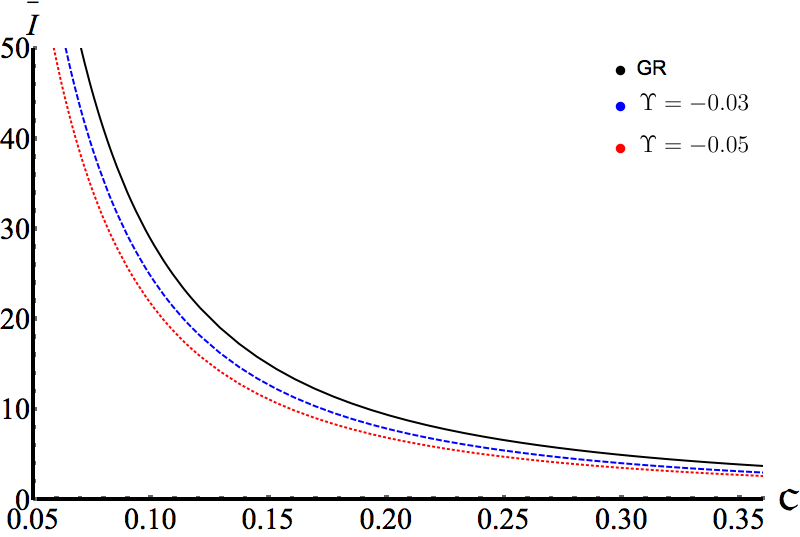}}
\caption{The $\bar{I}$--$\mathcal{C}$ relation in GR and Vainshtein breaking theories with parameters indicated in the figure. From top to bottom: GR (black, solid), $\Upsilon=-0.03$ (blue, dashed), $\Upsilon=-0.05$ (red, dotted). Figure taken from \cite{Sakstein:2016lyj}. }\label{fig:VBIC}
\end{figure}

A more robust method of testing gravity with relativistic stars is to use relations that are independent of the equation of state. In particular, it is well known in GR that there is a relation between the dimensionless moment of inertia $\bar{I}=I/(G^2M^3)$, where $I$ is the moment of inertia, and the compactness $\mathcal{C}=GM/R$ \cite{Breu:2016ufb,Yagi:2016bkt}. The compactness of a given star can be computed by solving the appropriate TOV equation but in order to compute the moment of inertia one needs to solve the equations for a slowly rotating star to first order in its angular velocity $\omega$. Given the complexity of the equations, we will once again sketch the procedure for calculating these quantities and refer the reader to \cite{Sakstein:2016lyj} for the full details. The method essentially follows the method of Hartle \cite{Hartle:1967he} applied to scalar-tensor theories.

The first step is to perturb the Schwarzchild-de Sitter metric to include the star's rotation with an angular velocity $\Omega$\footnote{The perturbation decays at infinity so that there is no change to the asymptotic spacetime.}. For slowly rotating objects, this plays the r\^{o}le of the small perturbation parameter. The quantity that must be calculated is $\omega(r)$, the coordinate angular velocity of the star as measured by a freely-falling observer.
One finds that the scalar is only perturbed at $\mathcal{O}(\Omega^2)$ whereas the $\mathcal{O}(\Omega)$ contribution to the perturbed tensor equations yield and equation of motion for $\omega$ of the form
\begin{equation}\label{eq:omeq}
\omega''=K_1(\delta\nu,\,\delta\lambda,\,\rho,\,\rho,\,P,\,P',\,\Upsilon)\omega'+K_0(\delta\nu,\,\delta\lambda,\,\rho,\,\rho,\,P,\,P',\,\Upsilon)\omega,
\end{equation}
where $K_1$ and $K_0$ are given in \cite{Sakstein:2016lyj} and reduce to their GR values of $K_1=4/r$ and $K_0=0$ when $\Upsilon=0$. The moment of inertia is found using the relation
\begin{equation}
\omega(R)=\Omega\left(1-\frac{G I}{R^3}\right).
\end{equation}

Reference \cite{Breu:2016ufb} found a $\bar{I}$--$\mathcal{C}$ relation of the form
\begin{equation}\label{IC}
\bar{I}=a_1\mathcal{C}^{-1}+a_2\mathcal{C}^{-2}+a_3\mathcal{C}^{-3}+a_1\mathcal{C}^{-4}
\end{equation}
for the GR relation well and so reference \cite{Sakstein:2016lyj} fit the Vainshtein breaking relation to the same function (the reader is referred to the original reference for the numerical coefficients). Their results are shown in figure \ref{fig:VBIC}. One can see that Vainshtein breaking theories also predict an $\bar{I}$--$\mathcal{C}$ that depends on $\Upsilon$ and furthermore that it is distinct from the GR prediction. Therefore, in principle, measuring the $\bar{I}$--$\mathcal{C}$ relation could place new bounds on Vainshtein breaking theories. In practice, this measurement is a while away since one needs to find highly relativistic systems where the (post-Newtonian) spin-orbit contribution to the precession can be measured. There are few known systems at the present time, although the next generation of radio surveys should be able to find more, making this measurement possible on a time-scale of a decade or so. Since Vainshtein breaking theories screen outside bodies, the measurement of the spin-orbit coupling, and therefore $I$ itself is the same as in GR.

\section{Astrophysical Tests of Couplings to Photons}

There have been many studies of chameleon theories that couple to photons via a term
\begin{equation}\label{eq:pc}
\Delta\mathcal{L}= \beta_\gamma\frac{\phi}{\mpl}F_\nm F^\nm=\frac{\phi}{M_\gamma}F_\nm F^\nm
\end{equation}
in the Lagrangian \cite{Brax:2010uq,Burrage:2016bwy,Burrage:2017qrf}. Mixing between the scalar and photons can induce both linear and circular polarizations into the (nominally unpolarized) starlight in the inter-galactic medium (IGM) \cite{Burrage:2008ii}. The lack of any observed polarization places the bound $M_\gamma>1.1\times 10^9 GeV$ provided $m_{\rm eff}(\phi_{\rm IGM})<1.1\times10^{-11}$ eV. The coupling \eqref{eq:pc} can also act as a loss mechanism whereby photons are converted to chameleons. This can result in deviations in the X-ray luminosity functions of active galactic nuclei (AGN), the lack of which places imposes $M_\gamma>10^{11}$ GeV for $m_{\rm eff}(\phi_{\rm AGN})<1.1\times10^{-12}$ eV. Similarly, the lack of any observed delpetion of CMB photons in the COMA cluster constrains $M_\gamma>1.1\times10^{9}$ GeV \cite{Davis:2010nj}. Finally, the depeletion of CMB photons increases the opacity of the universe and alters the distance-dulaity relation \cite{Avgoustidis:2010ju}, although current constraints are not competitive with those discussed above. This may change with data releases from current and next generation cosmological surveys.

\inputencoding{latin2}
\bibliography{ref}

\begin{thebibliography}{100}

\bibitem{Khoury:2003aq}
J.~Khoury and A.~Weltman, {\em Phys. Rev. Lett.} {\bf 93}  (2004)   171104,
  \href{http://arxiv.org/abs/astro-ph/0309300}{{\ttfamily
  arXiv:astro-ph/0309300 [astro-ph]}}.

\bibitem{Khoury:2003rn}
J.~Khoury and A.~Weltman, {\em Phys. Rev.} {\bf D69}  (2004)   044026,
  \href{http://arxiv.org/abs/astro-ph/0309411}{{\ttfamily
  arXiv:astro-ph/0309411 [astro-ph]}}.

\bibitem{Hinterbichler:2010es}
K.~Hinterbichler and J.~Khoury, {\em Phys. Rev. Lett.} {\bf 104}  (2010)
  231301, \href{http://arxiv.org/abs/1001.4525}{{\ttfamily arXiv:1001.4525
  [hep-th]}}.

\bibitem{Brax:2010gi}
P.~Brax, C.~van~de Bruck, A.-C. Davis and D.~Shaw, {\em Phys. Rev.} {\bf D82}
  (2010)   063519, \href{http://arxiv.org/abs/1005.3735}{{\ttfamily
  arXiv:1005.3735 [astro-ph.CO]}}.

\bibitem{Cabre:2012tq}
A.~Cabre, V.~Vikram, G.-B. Zhao, B.~Jain and K.~Koyama, {\em JCAP} {\bf 1207}
  (2012)   034, \href{http://arxiv.org/abs/1204.6046}{{\ttfamily
  arXiv:1204.6046 [astro-ph.CO]}}.

\bibitem{Zhao:2011cu}
G.-B. Zhao, B.~Li and K.~Koyama, {\em Phys. Rev. Lett.} {\bf 107}  (2011)
  071303, \href{http://arxiv.org/abs/1105.0922}{{\ttfamily arXiv:1105.0922
  [astro-ph.CO]}}.

\bibitem{Desmond:2017ctk}
H.~Desmond, P.~G. Ferreira, G.~Lavaux and J.~Jasche, {\em Mon. Not. Roy.
  Astron. Soc.} {\bf 474}  (2018) 3152,
  \href{http://arxiv.org/abs/1705.02420}{{\ttfamily arXiv:1705.02420
  [astro-ph.GA]}}.

\bibitem{Horndeski:1974wa}
G.~W. Horndeski, {\em Int. J. Theor. Phys.} {\bf 10}  (1974) 363.

\bibitem{Nicolis:2008in}
A.~Nicolis, R.~Rattazzi and E.~Trincherini, {\em Phys. Rev.} {\bf D79}  (2009)
   064036, \href{http://arxiv.org/abs/0811.2197}{{\ttfamily arXiv:0811.2197
  [hep-th]}}.

\bibitem{Gleyzes:2014dya}
J.~Gleyzes, D.~Langlois, F.~Piazza and F.~Vernizzi, {\em Phys. Rev. Lett.} {\bf
  114}  (2015)   211101, \href{http://arxiv.org/abs/1404.6495}{{\ttfamily
  arXiv:1404.6495 [hep-th]}}.

\bibitem{Gleyzes:2014qga}
J.~Gleyzes, D.~Langlois, F.~Piazza and F.~Vernizzi, {\em JCAP} {\bf 1502}
  (2015)   018, \href{http://arxiv.org/abs/1408.1952}{{\ttfamily
  arXiv:1408.1952 [astro-ph.CO]}}.

\bibitem{Langlois:2017mdk}
D.~Langlois, { {Degenerate Higher-Order Scalar-Tensor (DHOST) theories}}, in
  {\em {Proceedings, 52nd Rencontres de Moriond on Gravitation (Moriond
  Gravitation 2017): La Thuile, Italy, March 25-April 1, 2017}\/},  (2017), pp.
  221--228.
\newblock \href{http://arxiv.org/abs/1707.03625}{{\ttfamily arXiv:1707.03625
  [gr-qc]}}.

\bibitem{Kobayashi:2014ida}
T.~Kobayashi, Y.~Watanabe and D.~Yamauchi, {\em Phys. Rev.} {\bf D91}  (2015)
  064013, \href{http://arxiv.org/abs/1411.4130}{{\ttfamily arXiv:1411.4130
  [gr-qc]}}.

\bibitem{Koyama:2015oma}
K.~Koyama and J.~Sakstein, {\em Phys. Rev.} {\bf D91}  (2015)   124066,
  \href{http://arxiv.org/abs/1502.06872}{{\ttfamily arXiv:1502.06872
  [astro-ph.CO]}}.

\bibitem{Crisostomi:2017lbg}
M.~Crisostomi and K.~Koyama, {\em Phys. Rev.} {\bf D97}  (2018)   021301,
  \href{http://arxiv.org/abs/1711.06661}{{\ttfamily arXiv:1711.06661
  [astro-ph.CO]}}.

\bibitem{Sakstein:2017xjx}
J.~Sakstein and B.~Jain, {\em Phys. Rev. Lett.} {\bf 119}  (2017)   251303,
  \href{http://arxiv.org/abs/1710.05893}{{\ttfamily arXiv:1710.05893
  [astro-ph.CO]}}.

\bibitem{Langlois:2017dyl}
D.~Langlois, R.~Saito, D.~Yamauchi and K.~Noui  (2017)
  \href{http://arxiv.org/abs/1711.07403}{{\ttfamily arXiv:1711.07403 [gr-qc]}}.

\bibitem{Dima:2017pwp}
A.~Dima and F.~Vernizzi  (2017)
  \href{http://arxiv.org/abs/1712.04731}{{\ttfamily arXiv:1712.04731 [gr-qc]}}.

\bibitem{Bellini:2014fua}
E.~Bellini and I.~Sawicki, {\em JCAP} {\bf 1407}  (2014)   050,
  \href{http://arxiv.org/abs/1404.3713}{{\ttfamily arXiv:1404.3713
  [astro-ph.CO]}}.

\bibitem{Hui:2009kc}
L.~Hui, A.~Nicolis and C.~Stubbs, {\em Phys. Rev.} {\bf D80}  (2009)   104002,
  \href{http://arxiv.org/abs/0905.2966}{{\ttfamily arXiv:0905.2966
  [astro-ph.CO]}}.

\bibitem{Mota:2006ed}
D.~F. Mota and D.~J. Shaw, {\em Phys. Rev. Lett.} {\bf 97}  (2006)   151102,
  \href{http://arxiv.org/abs/hep-ph/0606204}{{\ttfamily arXiv:hep-ph/0606204
  [hep-ph]}}.

\bibitem{Mota:2006fz}
D.~F. Mota and D.~J. Shaw, {\em Phys. Rev.} {\bf D75}  (2007)   063501,
  \href{http://arxiv.org/abs/hep-ph/0608078}{{\ttfamily arXiv:hep-ph/0608078
  [hep-ph]}}.

\bibitem{Hiramatsu:2012xj}
T.~Hiramatsu, W.~Hu, K.~Koyama and F.~Schmidt, {\em Phys. Rev.} {\bf D87}
  (2013)   063525, \href{http://arxiv.org/abs/1209.3364}{{\ttfamily
  arXiv:1209.3364 [hep-th]}}.

\bibitem{Bloomfield:2014zfa}
J.~K. Bloomfield, C.~Burrage and A.-C. Davis, {\em Phys. Rev.} {\bf D91}
  (2015)   083510, \href{http://arxiv.org/abs/1408.4759}{{\ttfamily
  arXiv:1408.4759 [gr-qc]}}.

\bibitem{Brito:2014ifa}
R.~Brito, A.~Terrana, M.~Johnson and V.~Cardoso, {\em Phys. Rev.} {\bf D90}
  (2014)   124035, \href{http://arxiv.org/abs/1409.0886}{{\ttfamily
  arXiv:1409.0886 [hep-th]}}.

\bibitem{Andrews:2013qva}
M.~Andrews, Y.-Z. Chu and M.~Trodden, {\em Phys. Rev.} {\bf D88}  (2013)
  084028, \href{http://arxiv.org/abs/1305.2194}{{\ttfamily arXiv:1305.2194
  [astro-ph.CO]}}.

\bibitem{Avilez-Lopez:2015dja}
A.~Avilez-Lopez, A.~Padilla, P.~M. Saffin and C.~Skordis, {\em JCAP} {\bf 1506}
   (2015)   044, \href{http://arxiv.org/abs/1501.01985}{{\ttfamily
  arXiv:1501.01985 [gr-qc]}}.

\bibitem{Kanti:1995vq}
P.~Kanti, N.~E. Mavromatos, J.~Rizos, K.~Tamvakis and E.~Winstanley, {\em Phys.
  Rev.} {\bf D54}  (1996) 5049,
  \href{http://arxiv.org/abs/hep-th/9511071}{{\ttfamily arXiv:hep-th/9511071
  [hep-th]}}.

\bibitem{Sotiriou:2013qea}
T.~P. Sotiriou and S.-Y. Zhou, {\em Phys. Rev. Lett.} {\bf 112}  (2014)
  251102, \href{http://arxiv.org/abs/1312.3622}{{\ttfamily arXiv:1312.3622
  [gr-qc]}}.

\bibitem{Sotiriou:2014pfa}
T.~P. Sotiriou and S.-Y. Zhou, {\em Phys. Rev.} {\bf D90}  (2014)   124063,
  \href{http://arxiv.org/abs/1408.1698}{{\ttfamily arXiv:1408.1698 [gr-qc]}}.

\bibitem{Dong:2017toi}
R.~Dong, J.~Sakstein and D.~Stojkovic, {\em Phys. Rev.} {\bf D96}  (2017)
  064048, \href{http://arxiv.org/abs/1709.01641}{{\ttfamily arXiv:1709.01641
  [gr-qc]}}.

\bibitem{Silva:2017uqg}
H.~O. Silva, J.~Sakstein, L.~Gualtieri, T.~P. Sotiriou and E.~Berti  (2017)
  \href{http://arxiv.org/abs/1711.02080}{{\ttfamily arXiv:1711.02080 [gr-qc]}}.

\bibitem{Hui:2012qt}
L.~Hui and A.~Nicolis, {\em Phys. Rev. Lett.} {\bf 110}  (2013)   241104,
  \href{http://arxiv.org/abs/1202.1296}{{\ttfamily arXiv:1202.1296 [hep-th]}}.

\bibitem{Davis:2014tea}
A.-C. Davis, R.~Gregory, R.~Jha and J.~Muir, {\em JCAP} {\bf 1408}  (2014)
  033, \href{http://arxiv.org/abs/1402.4737}{{\ttfamily arXiv:1402.4737
  [astro-ph.CO]}}.

\bibitem{Davis:2016avf}
A.-C. Davis, R.~Gregory and R.~Jha, {\em JCAP} {\bf 1610}  (2016)   024,
  \href{http://arxiv.org/abs/1607.08607}{{\ttfamily arXiv:1607.08607 [gr-qc]}}.

\bibitem{Kippenhahn1994}
R.~{Kippenhahn} and A.~{Weigert}, {\em {Stellar Structure and Evolution}} 1994.

\bibitem{Chang:2010xh}
P.~Chang and L.~Hui, {\em Astrophys. J.} {\bf 732}  (2011)  ~25,
  \href{http://arxiv.org/abs/1011.4107}{{\ttfamily arXiv:1011.4107
  [astro-ph.CO]}}.

\bibitem{Davis:2011qf}
A.-C. Davis, E.~A. Lim, J.~Sakstein and D.~Shaw, {\em Phys. Rev.} {\bf D85}
  (2012)   123006, \href{http://arxiv.org/abs/1102.5278}{{\ttfamily
  arXiv:1102.5278 [astro-ph.CO]}}.

\bibitem{Jain:2012tn}
B.~Jain, V.~Vikram and J.~Sakstein, {\em Astrophys. J.} {\bf 779}  (2013)  ~39,
  \href{http://arxiv.org/abs/1204.6044}{{\ttfamily arXiv:1204.6044
  [astro-ph.CO]}}.

\bibitem{Sakstein:2013pda}
J.~Sakstein, {\em Phys. Rev.} {\bf D88}  (2013)   124013,
  \href{http://arxiv.org/abs/1309.0495}{{\ttfamily arXiv:1309.0495
  [astro-ph.CO]}}.

\bibitem{Sakstein:2015oqa}
J.~Sakstein, {Astrophysical Tests of Modified Gravity}, PhD thesis, Cambridge
  U., DAMTP  (2014).

\bibitem{Sakstein:2015aqx}
J.~Sakstein and K.~Koyama, {\em Int. J. Mod. Phys.} {\bf D24}  (2015)
  1544021.

\bibitem{Saito:2015fza}
R.~Saito, D.~Yamauchi, S.~Mizuno, J.~Gleyzes and D.~Langlois, {\em JCAP} {\bf
  1506}  (2015)   008, \href{http://arxiv.org/abs/1503.01448}{{\ttfamily
  arXiv:1503.01448 [gr-qc]}}.

\bibitem{Jain:2015edg}
R.~K. Jain, C.~Kouvaris and N.~G. Nielsen, {\em Phys. Rev. Lett.} {\bf 116}
  (2016)   151103, \href{http://arxiv.org/abs/1512.05946}{{\ttfamily
  arXiv:1512.05946 [astro-ph.CO]}}.

\bibitem{Chandrasekhar1939}
S.~{Chandrasekhar}, {\em {An introduction to the study of stellar structure}}
  1939.

\bibitem{Horedt1987}
G.~P. {Horedt}, {\em \aap} {\bf 177} (May 1987) 117.

\bibitem{Horedt2004}
G.~P. {Horedt} (ed.), {\em {Polytropes - Applications in Astrophysics and
  Related Fields}}, Astrophysics and Space Science Library Vol.~306, July 2004.

\bibitem{Paxton:2010ji}
B.~Paxton, L.~Bildsten, A.~Dotter, F.~Herwig, P.~Lesaffre and F.~Timmes, {\em
  Astrophys. J. Suppl.} {\bf 192}  (2011)  ~3,
  \href{http://arxiv.org/abs/1009.1622}{{\ttfamily arXiv:1009.1622
  [astro-ph.SR]}}.

\bibitem{Paxton:2013pj}
B.~Paxton {\em et~al.}, {\em Astrophys. J. Suppl.} {\bf 208}  (2013)  ~4,
  \href{http://arxiv.org/abs/1301.0319}{{\ttfamily arXiv:1301.0319
  [astro-ph.SR]}}.

\bibitem{Paxton:2015jva}
B.~Paxton {\em et~al.}, {\em Astrophys. J. Suppl.} {\bf 220}  (2015)  ~15,
  \href{http://arxiv.org/abs/1506.03146}{{\ttfamily arXiv:1506.03146
  [astro-ph.SR]}}.

\bibitem{Najafi:2018apy}
S.~Najafi, M.~T. Mirtorabi, Z.~Ansari and D.~F. Mota  (2018)
  \href{http://arxiv.org/abs/1802.04001}{{\ttfamily arXiv:1802.04001 [gr-qc]}}.

\bibitem{Cox1980}
J.~P. {Cox}, {\em {Theory of stellar pulsation}} 1980.

\bibitem{Silvestri:2011ch}
A.~Silvestri, {\em Phys. Rev. Lett.} {\bf 106}  (2011)   251101,
  \href{http://arxiv.org/abs/1103.4013}{{\ttfamily arXiv:1103.4013
  [astro-ph.CO]}}.

\bibitem{Upadhye:2013nfa}
A.~Upadhye and J.~H. Steffen  (2013)
  \href{http://arxiv.org/abs/1306.6113}{{\ttfamily arXiv:1306.6113
  [astro-ph.CO]}}.

\bibitem{Brax:2013uh}
P.~Brax, A.-C. Davis and J.~Sakstein, {\em Class. Quant. Grav.} {\bf 31}
  (2014)   225001, \href{http://arxiv.org/abs/1301.5587}{{\ttfamily
  arXiv:1301.5587 [gr-qc]}}.

\bibitem{Chandrasekhar1964}
S.~{Chandrasekhar}, {\em \apj} {\bf 140} (August 1964)   417.

\bibitem{Sakstein:2016lyj}
J.~Sakstein, M.~Kenna-Allison and K.~Koyama, {\em JCAP} {\bf 1703}  (2017)
  007, \href{http://arxiv.org/abs/1611.01062}{{\ttfamily arXiv:1611.01062
  [gr-qc]}}.

\bibitem{Delgaty:1998uy}
M.~S.~R. Delgaty and K.~Lake, {\em Comput. Phys. Commun.} {\bf 115}  (1998)
  395, \href{http://arxiv.org/abs/gr-qc/9809013}{{\ttfamily arXiv:gr-qc/9809013
  [gr-qc]}}.

\bibitem{Burrows:1992fg}
A.~Burrows and J.~Liebert, {\em Rev. Mod. Phys.} {\bf 65}  (1993) 301.

\bibitem{Sakstein:2015zoa}
J.~Sakstein, {\em Phys. Rev. Lett.} {\bf 115}  (2015)   201101,
  \href{http://arxiv.org/abs/1510.05964}{{\ttfamily arXiv:1510.05964
  [astro-ph.CO]}}.

\bibitem{Sakstein:2015aac}
J.~Sakstein, {\em Phys. Rev.} {\bf D92}  (2015)   124045,
  \href{http://arxiv.org/abs/1511.01685}{{\ttfamily arXiv:1511.01685
  [astro-ph.CO]}}.

\bibitem{Saumon1995}
D.~{Saumon}, G.~{Chabrier} and H.~M. {van Horn}, {\em \apjs} {\bf 99} (August
  1995)   713.

\bibitem{Chabrier:2008bc}
G.~Chabrier, I.~Baraffe, J.~Leconte, J.~Gallardo and T.~barman, {\em AIP Conf.
  Proc.} {\bf 1094}  (2009) 102,
  \href{http://arxiv.org/abs/0810.5085}{{\ttfamily arXiv:0810.5085
  [astro-ph]}}.

\bibitem{Segransan:2000jq}
D.~Segransan, X.~Delfosse, T.~Forveille, J.~L. Beuzit, S.~Udry, C.~Perrier and
  M.~Mayor, {\em Astron. Astrophys.} {\bf 364}  (2000)   665,
  \href{http://arxiv.org/abs/astro-ph/0010585}{{\ttfamily
  arXiv:astro-ph/0010585 [astro-ph]}}.

\bibitem{Salpeter1955}
E.~E. {Salpeter}, {\em \apj} {\bf 121} (January 1955)   161.

\bibitem{Kippenhahn1970}
R.~{Kippenhahn}, {\em \aap} {\bf 8} (September 1970)  ~50.

\bibitem{Henry:1993tk}
T.~J. Henry and D.~W. McCarthy, Jr., {\em Astron. J.} {\bf 106}  (1993) 773.

\bibitem{Shapiro1983}
S.~L. {Shapiro} and S.~A. {Teukolsky}, {\em {Black holes, white dwarfs, and
  neutron stars: The physics of compact objects}} 1983.

\bibitem{Holberg:2012pu}
J.~B. Holberg, T.~D. Oswalt and M.~A. Barstow, {\em Astron. J.} {\bf 143}
  (2012)  ~68, \href{http://arxiv.org/abs/1201.3822}{{\ttfamily arXiv:1201.3822
  [astro-ph.SR]}}.

\bibitem{Hachisu:2000nx}
I.~Hachisu and M.~Kato, {\em Astrophys. J.} {\bf 540}  (2000)   447,
  \href{http://arxiv.org/abs/astro-ph/0003471}{{\ttfamily
  arXiv:astro-ph/0003471 [astro-ph]}}.

\bibitem{Mereghetti:2011bh}
S.~Mereghetti, N.~La~Palombara, A.~Tiengo, F.~Pizzolato, P.~Esposito, P.~A.
  Woudt, G.~L. Israel and L.~Stella, {\em Astrophys. J.} {\bf 737}  (2011)
  ~51, \href{http://arxiv.org/abs/1105.6227}{{\ttfamily arXiv:1105.6227
  [astro-ph.HE]}}.

\bibitem{Freedman:2010xv}
W.~L. Freedman and B.~F. Madore, {\em Ann. Rev. Astron. Astrophys.} {\bf 48}
  (2010) 673, \href{http://arxiv.org/abs/1004.1856}{{\ttfamily arXiv:1004.1856
  [astro-ph.CO]}}.

\bibitem{Jain:2011ji}
B.~Jain and J.~VanderPlas, {\em JCAP} {\bf 1110}  (2011)   032,
  \href{http://arxiv.org/abs/1106.0065}{{\ttfamily arXiv:1106.0065
  [astro-ph.CO]}}.

\bibitem{Navarro:1995iw}
J.~F. Navarro, C.~S. Frenk and S.~D.~M. White, {\em Astrophys. J.} {\bf 462}
  (1996) 563, \href{http://arxiv.org/abs/astro-ph/9508025}{{\ttfamily
  arXiv:astro-ph/9508025 [astro-ph]}}.

\bibitem{Swaters2011}
R.~A. {Swaters}, R.~{Sancisi}, T.~S. {van Albada} and J.~M. {van der Hulst},
  {\em \apj} {\bf 729} (March 2011)   118,
  \href{http://arxiv.org/abs/1101.3120}{{\ttfamily arXiv:1101.3120}}.

\bibitem{Vikram:2014uza}
V.~Vikram, J.~Sakstein, C.~Davis and A.~Neil  (2014)
  \href{http://arxiv.org/abs/1407.6044}{{\ttfamily arXiv:1407.6044
  [astro-ph.CO]}}.

\bibitem{Vikram:2013uba}
V.~Vikram, A.~Cabré, B.~Jain and J.~T. VanderPlas, {\em JCAP} {\bf 1308}
  (2013)   020, \href{http://arxiv.org/abs/1303.0295}{{\ttfamily
  arXiv:1303.0295 [astro-ph.CO]}}.

\bibitem{Desmond:2018euk}
H.~Desmond, P.~G. Ferreira, G.~Lavaux and J.~Jasche  (2018)
  \href{http://arxiv.org/abs/1802.07206}{{\ttfamily arXiv:1802.07206
  [astro-ph.CO]}}.

\bibitem{Burrage:2016bwy}
C.~Burrage and J.~Sakstein, {\em JCAP} {\bf 1611}  (2016)   045,
  \href{http://arxiv.org/abs/1609.01192}{{\ttfamily arXiv:1609.01192
  [astro-ph.CO]}}.

\bibitem{Burrage:2017qrf}
C.~Burrage and J.~Sakstein, {\em Living Rev. Rel.} {\bf 21}  (2018)  ~1,
  \href{http://arxiv.org/abs/1709.09071}{{\ttfamily arXiv:1709.09071
  [astro-ph.CO]}}.

\bibitem{Wilcox:2016guw}
H.~Wilcox, R.~C. Nichol, G.-B. Zhao, D.~Bacon, K.~Koyama and A.~K. Romer, {\em
  Mon. Not. Roy. Astron. Soc.} {\bf 462}  (2016) 715,
  \href{http://arxiv.org/abs/1603.05911}{{\ttfamily arXiv:1603.05911
  [astro-ph.CO]}}.

\bibitem{Terukina:2013eqa}
A.~Terukina, L.~Lombriser, K.~Yamamoto, D.~Bacon, K.~Koyama and R.~C. Nichol,
  {\em JCAP} {\bf 1404}  (2014)   013,
  \href{http://arxiv.org/abs/1312.5083}{{\ttfamily arXiv:1312.5083
  [astro-ph.CO]}}.

\bibitem{Wilcox:2015kna}
H.~Wilcox {\em et~al.}, {\em Mon. Not. Roy. Astron. Soc.} {\bf 452}  (2015)
  1171, \href{http://arxiv.org/abs/1504.03937}{{\ttfamily arXiv:1504.03937
  [astro-ph.CO]}}.

\bibitem{Sakstein:2016ggl}
J.~Sakstein, H.~Wilcox, D.~Bacon, K.~Koyama and R.~C. Nichol, {\em JCAP} {\bf
  1607}  (2016)   019, \href{http://arxiv.org/abs/1603.06368}{{\ttfamily
  arXiv:1603.06368 [astro-ph.CO]}}.

\bibitem{Sakstein:2017bws}
J.~Sakstein, B.~Jain, J.~S. Heyl and L.~Hui, {\em Astrophys. J.} {\bf 844}
  (2017)   L14, \href{http://arxiv.org/abs/1704.02425}{{\ttfamily
  arXiv:1704.02425 [astro-ph.CO]}}.

\bibitem{Dvali:2002vf}
G.~Dvali, A.~Gruzinov and M.~Zaldarriaga, {\em Phys. Rev.} {\bf D68}  (2003)
  024012, \href{http://arxiv.org/abs/hep-ph/0212069}{{\ttfamily
  arXiv:hep-ph/0212069 [hep-ph]}}.

\bibitem{Sakstein:2017pqi}
J.~Sakstein  (2017) \href{http://arxiv.org/abs/1710.03156}{{\ttfamily
  arXiv:1710.03156 [astro-ph.CO]}}.

\bibitem{Hui:2012jb}
L.~Hui and A.~Nicolis, {\em Phys. Rev. Lett.} {\bf 109}  (2012)   051304,
  \href{http://arxiv.org/abs/1201.1508}{{\ttfamily arXiv:1201.1508
  [astro-ph.CO]}}.

\bibitem{Schmidt:2010jr}
F.~Schmidt, {\em Phys. Rev.} {\bf D81}  (2010)   103002,
  \href{http://arxiv.org/abs/1003.0409}{{\ttfamily arXiv:1003.0409
  [astro-ph.CO]}}.

\bibitem{Diemer:2014xya}
B.~Diemer and A.~V. Kravtsov, {\em Astrophys. J.} {\bf 789}  (2014)  ~1,
  \href{http://arxiv.org/abs/1401.1216}{{\ttfamily arXiv:1401.1216
  [astro-ph.CO]}}.

\bibitem{Asvathaman:2015nna}
A.~Asvathaman, J.~S. Heyl and L.~Hui, {\em Mon. Not. Roy. Astron. Soc.} {\bf
  465}  (2017) 3261, \href{http://arxiv.org/abs/1506.07607}{{\ttfamily
  arXiv:1506.07607 [astro-ph.GA]}}.

\bibitem{Brax:2017wcj}
P.~Brax, A.-C. Davis and R.~Jha, {\em Phys. Rev.} {\bf D95}  (2017)   083514,
  \href{http://arxiv.org/abs/1702.02983}{{\ttfamily arXiv:1702.02983 [gr-qc]}}.

\bibitem{Chagoya:2014fza}
J.~Chagoya, K.~Koyama, G.~Niz and G.~Tasinato, {\em JCAP} {\bf 1410}  (2014)
  055, \href{http://arxiv.org/abs/1407.7744}{{\ttfamily arXiv:1407.7744
  [hep-th]}}.

\bibitem{Berti:2015itd}
E.~Berti {\em et~al.}, {\em Class. Quant. Grav.} {\bf 32}  (2015)   243001,
  \href{http://arxiv.org/abs/1501.07274}{{\ttfamily arXiv:1501.07274 [gr-qc]}}.

\bibitem{Babichev:2016rlq}
E.~Babichev, C.~Charmousis and A.~Lehébel, {\em Class. Quant. Grav.} {\bf 33}
  (2016)   154002, \href{http://arxiv.org/abs/1604.06402}{{\ttfamily
  arXiv:1604.06402 [gr-qc]}}.

\bibitem{Babichev:2009td}
E.~Babichev and D.~Langlois, {\em Phys. Rev.} {\bf D80}  (2009)   121501,
  \href{http://arxiv.org/abs/0904.1382}{{\ttfamily arXiv:0904.1382 [gr-qc]}},
  [Erratum: Phys. Rev.D81,069901(2010)].

\bibitem{Babichev:2009fi}
E.~Babichev and D.~Langlois, {\em Phys. Rev.} {\bf D81}  (2010)   124051,
  \href{http://arxiv.org/abs/0911.1297}{{\ttfamily arXiv:0911.1297 [gr-qc]}}.

\bibitem{Babichev:2016jom}
E.~Babichev, K.~Koyama, D.~Langlois, R.~Saito and J.~Sakstein, {\em Class.
  Quant. Grav.} {\bf 33}  (2016)   235014,
  \href{http://arxiv.org/abs/1606.06627}{{\ttfamily arXiv:1606.06627 [gr-qc]}}.

\bibitem{Sakstein:2016oel}
J.~Sakstein, E.~Babichev, K.~Koyama, D.~Langlois and R.~Saito, {\em Phys. Rev.}
  {\bf D95}  (2017)   064013, \href{http://arxiv.org/abs/1612.04263}{{\ttfamily
  arXiv:1612.04263 [gr-qc]}}.

\bibitem{Demorest:2010bx}
P.~Demorest, T.~Pennucci, S.~Ransom, M.~Roberts and J.~Hessels, {\em Nature}
  {\bf 467}  (2010) 1081, \href{http://arxiv.org/abs/1010.5788}{{\ttfamily
  arXiv:1010.5788 [astro-ph.HE]}}.

\bibitem{Breu:2016ufb}
C.~Breu and L.~Rezzolla, {\em Mon. Not. Roy. Astron. Soc.} {\bf 459}  (2016)
  646, \href{http://arxiv.org/abs/1601.06083}{{\ttfamily arXiv:1601.06083
  [gr-qc]}}.

\bibitem{Yagi:2016bkt}
K.~Yagi and N.~Yunes, {\em Phys. Rept.} {\bf 681}  (2017) 1,
  \href{http://arxiv.org/abs/1608.02582}{{\ttfamily arXiv:1608.02582 [gr-qc]}}.

\bibitem{Hartle:1967he}
J.~B. Hartle, {\em Astrophys. J.} {\bf 150}  (1967) 1005.

\bibitem{Brax:2010uq}
P.~Brax, C.~Burrage, A.-C. Davis, D.~Seery and A.~Weltman, {\em Phys. Lett.}
  {\bf B699}  (2011) 5, \href{http://arxiv.org/abs/1010.4536}{{\ttfamily
  arXiv:1010.4536 [hep-th]}}.

\bibitem{Burrage:2008ii}
C.~Burrage, A.-C. Davis and D.~J. Shaw, {\em Phys. Rev.} {\bf D79}  (2009)
  044028, \href{http://arxiv.org/abs/0809.1763}{{\ttfamily arXiv:0809.1763
  [astro-ph]}}.

\bibitem{Davis:2010nj}
A.-C. Davis, C.~A.~O. Schelpe and D.~J. Shaw, {\em Phys. Rev.} {\bf D83}
  (2011)   044006, \href{http://arxiv.org/abs/1008.1880}{{\ttfamily
  arXiv:1008.1880 [astro-ph.CO]}}.

\bibitem{Avgoustidis:2010ju}
A.~Avgoustidis, C.~Burrage, J.~Redondo, L.~Verde and R.~Jimenez, {\em JCAP}
  {\bf 1010}  (2010)   024, \href{http://arxiv.org/abs/1004.2053}{{\ttfamily
  arXiv:1004.2053 [astro-ph.CO]}}.

\end{thebibliography}
\inputencoding{utf8}
\bibliographystyle{ws-ijmpd}

\end{document}